\documentclass[12pt]{article}

\usepackage{amsfonts,amssymb,graphics,epsfig,verbatim,bm,latexsym,amsmath,url,amsbsy,cancel}
\usepackage{dsfont}
\usepackage{float}
\usepackage{multirow,natbib}
\usepackage[shortlabels]{enumitem}
\newtheorem{theorem}{Theorem}
\newtheorem{assumption}{Assumption}

\newtheorem{lemma}{Lemma}

\newtheorem{definition}{Definition}

\newcommand\independent{\protect\mathpalette{\protect\independenT}{\perp}}
\def\independenT#1#2{\mathrel{\rlap{$#1#2$}\mkern2mu{#1#2}}}
\DeclareMathOperator*{\argmin}{arg\,min}
\newcounter{bean}

\usepackage{color}
\newif\ifcorr

\begin{document}
\corrfalse

\title{Instrumental variable quantile regression under random right censoring}

\author{
{\large Jad B\textsc{eyhum}
\footnote{ORSTAT, KU Leuven. Financial support from the European Research Council (2016-2022, Horizon 2020 / ERC grant agreement No.\ 694409) is gratefully acknowledged.}}\\\texttt{\small jad.beyhum@gmail.com}
\and
\addtocounter{footnote}{2}
{\large Lorenzo T\textsc{edesco}} $^*$
\\\texttt{\small lorenzo.tedesco@kuleuven.be}
\and
{\large Ingrid V\textsc{{an} K{eilegom}} $^*$}
\\\texttt{\small ingrid.vankeilegom@kuleuven.be}
}

\maketitle


    \begin{abstract} This paper studies a semiparametric quantile regression model with endogenous variables and random right censoring. The endogeneity issue is solved using instrumental variables. It is assumed that the structural quantile of the logarithm of the outcome variable is linear in the covariates and censoring is independent. The regressors and instruments can be either continuous or discrete. The specification generates a continuum of equations of which the quantile regression coefficients are a solution. Identification is obtained when this system of equations has a unique solution. Our estimation procedure solves an empirical analogue of the system of equations.  We derive conditions under which the estimator is asymptotically normal and prove the validity of a bootstrap procedure for inference. The finite sample performance of the approach is evaluated through numerical simulations. An application to the national Job Training Partnership Act study illustrates the method.


\bigskip

\def\baselinestretch{1.3}

\newpage
\normalsize
    \end{abstract}


   \section{Introduction} \label{sec:introduction}

Let us consider a setting where the researcher is interested in the causal effect of some regressors $Z$ on a duration outcome $T$. We wish to recover the structural quantile of $T$ if the treatment were set to a particular value. This task is often complicated by two issues. The first one is endogeneity, that is the variable $Z$ and the unobserved heterogeneity $U$ are dependent. In this case, the causal effect of the treatment is not characterized by the conditional distribution of $T$ given $Z$. The second problem is right censoring. It happens, for instance, when the period of observation of the subjects in the dataset is limited. If the length of the follow-up does not depend on the subject's characteristics, the censoring is called independent.

In this paper, we propose a new instrumental variable estimator which allows treating both issues. The endogeneity of the treatment is addressed through an instrumental variable $W$, independent of the error term $U$ of the model but sufficiently related to the treatment. We assume that censoring is random and independent. The structural quantile of $\log T$ is linear in the variables, which makes the model semiparametric. The causal effect at a given quantile $u$ is therefore characterized by a finite dimensional parameter vector $\beta_0(u)$. This model allows to handle several continuous or discrete variables and is standard in the literature on censored quantile regression (see below for related literature). 

Specifically, we show that $\beta_0(u)$ is the solution to a continuum of integral equations. This result allows us to derive local and global identification results. Then, we propose a new minimum distance estimator solving an estimated version of the system of identification equations. {\ifcorr\color{red} \else \color{black} \fi We address censoring through a weighting scheme.} {\ifcorr\color{red} \else \color{black} \fi The estimator is asymptotically normal under some conditions}. {\ifcorr\color{red} \else \color{black} \fi We prove the validity of a bootstrap procedure for inference}. Finite {\ifcorr\color{red} \else \color{black} \fi sample} properties of the estimator are assessed through simulations. The procedure is illustrated {\ifcorr\color{red} \else \color{black} \fi through an} application to the national Job Training Partnership Act (JTPA) study. An appealing feature of our approach is that the instrument and the covariates can be either discrete or continuous. \\

\noindent\textbf{Related literature} 
 This paper is first related to the literature on censored quantile regression with exogenous regressors, which studies a model where the conditional quantile of $\log(T)$ is linear in the regressors without relying on instrumental variables. Earlier works, such as \citet{powell1984least,powell1986censored, khan2001two, fitzenberger199715}, proposed estimators for this model in the special case where the censoring time is constant and observed.   An alternative approach is \citet{buchinsky1998alternative} where the censoring time is an unknown function of the regressors but does not need to be always observed. \citet{honore2002quantile}  later allowed for random right censoring independent of the outcome variable and the regressors. Then, \citet{chernozhukov2002three} relaxed this condition by assuming only that the duration variable and the censoring time are independent conditional on the regressors. Both \citet{honore2002quantile} and \citet{chernozhukov2002three} required the censoring time to be observed. The approach developed in \citet{portnoy2003censored} relies on the same independence assumption as \citet{chernozhukov2002three} but allows the censoring time to be unobserved for uncensored observations. Alternative methods have been proposed in \citet{peng2008survival}, which is based on martingales, or in \citet{yang2018new}, which follows a data augmentation approach.
 The methods of \citet{portnoy2003censored}, \citet{peng2008survival} and \citet{yang2018new} rely on the global assumption that all the conditional quantiles of $\log(T)$ are linear in the regressors.  \citet{wang2009locally} relaxed this global assumption by assuming linearity only for the quantile of interest (the approach of the present paper would also work under a local linearity assumption of this type). Other estimators which only impose similar local conditions are \citet{de2019adapted}, which is based on adjusting the standard quantile loss function in order to accommodate randomly censored data, and \citet{de2020linear}, who propose a minimum-distance estimator. For more details about each of these estimators, we refer the reader to the recent review by \citet{Peng2021}.

There is also a large literature on instrumental variable approaches with randomly right censored duration outcomes, where it is not assumed that the quantile of $\log T$ is linear in the regressors. We only cite here some works for reasons of brevity. {\ifcorr\color{red} \else \color{black} \fi Some articles study different semiparametric models, see e.g. \citet{tchetgen2015instrumental} for an additive hazard model or \citet{martinussen2019instrumental} and \cite{wang2018learning} for the Cox model}. On the other hand, \citet{frandsen2015treatment}, \citet{sant2016program}, \citet{richardson2017nonparametric}, \citet{blanco2019bounds}, \citet{sant2021nonparametric}, \citet{beyhum2022nonparametric} have developed nonparametric approaches with categorical treatment and instrument, whereas \citet{centorrinoflorens2021} proposed a nonparametric estimator when the variables are continuous and the model is additive.

Finally, this paper is most related to the literature on instrumental variable methods under right censoring assuming that the structural quantile of $\log T$ is linear in the regressors. First, there are papers, such as \citet{blundell2007censored}, \citet{hong2003inference}, \citet{chen2020semiparametric} and \citet{wang2021moment}, assuming that the censoring time $C$ is constant. 
In contrast, the present article allows $C$ to be random. Such a distinction is particularly relevant in studies where the duration of follow-up depends on the date of entry in the study. Next, although censoring is random in \cite{chernozhukov2015quantile}, they assumed that $C$ is always observed, which we do not. Their approach is based on a control function requiring a separate specification for the relation between the regressors and the instrument. We do not need these additional structural assumptions. Also, they assume that the endogenous regressor is continuous, while our approach allows both discrete and continuous cases. {\ifcorr\color{red} \else \color{black} \fi \citet{khan2009inference} present an alternative estimator}. Their method is designed to handle dependent censoring. It requires a restrictive support conditions {\ifcorr\color{red} \else \color{black} \fi (see condition IV2, page 110 in \cite{khan2009inference})}, which may not hold when censoring is independent. We do not need such a support assumption. Moreover, \citet{wei2021estimation} studied the case where there are exogenous covariates, a single binary endogenous variable and a binary instrument.  They imposed a monotonicity assumption (as in \citet{angrist1996identification}), which means that the value of the treatment is increasing in the instrument. They identified and estimated the quantile treatment effects over the population of compliers, that is the subjects whose treatment status changes with the instrument. Instead, the present paper allows for nonbinary endogenous regressors and instruments. Additionally, thanks to a rank invariance assumption as in \citet{chernozhukov2005iv}, we identify the quantile treatment effects on the whole population rather than only on the compliers.\\

\noindent\textbf{Outline} The paper is organised as follows. The model specification is given in Section 2. Identification results are presented in Section 3. Section 4 is devoted to estimation and inference. Sections 5 and 6 present the simulations and the empirical application, respectively. All proofs can be found in the supplementary material. The code for the simulations and the empirical application is in the replication package.

\section{The model} \label{sec:2}
\setcounter{equation}{0}
\setcounter{theorem}{0}
\setcounter{definition}{0}
Denote by $T$ the duration outcome variable with values in $\mathbb{R}_+$, by $Z$ a random vector of regressors with support $\mathcal{Z}\subset \mathbb{R}^K$. Let also $T(z)$ be the potential outcome of $T$ under treatment $z\in \mathcal{Z}$. By consistency, it holds that $T=T(Z)$. The random variable $U$ is the unobserved heterogeneity of the model. We normalize it to follow a uniform distribution on the interval $[0,1]$. We suppose that there exists quantile linear regression coefficients $\beta_0(\cdot):(0,1) \rightarrow\mathbb{R}^K$ such that the following relationship among the potential outcomes holds:
\begin{equation}    \label{eq:model}
     T(z) = \exp(z^\top\beta_0(U)),\ \textrm{for all}\ z\in\mathcal{Z},\ \textrm{a.s.}
\end{equation}
We also assume that the mapping $\beta_0(\cdot)$ is continuously differentiable everywhere with derivative denoted by $D\beta_0(\cdot)$ and for all $u\in (0,1)$ and $z \in \mathcal{Z}$, $z^\top D\beta_0(u)>0$. This ensures that $z^\top\beta_0(\cdot)$ is a well-behaved quantile function, in the sense that $u\in(0,1)\mapsto z^\top \beta_0(u)$ is strictly increasing. 
Under this last condition, for any value $u\in(0,1)$, the $u$-quantile of $\log T(z)$ is equal to $z^{\top}\beta_0(u)$, and $\beta_0(u)$ measures the causal effect of $Z$ on this structural $u$-quantile. Our goal is to identify and estimate $\beta_0(u)$, for some given $u\in(0,1)$. Note that we do not define $\beta_0(\cdot)$ on $\{0,1\}$. This choice avoids pathological behaviours happening at the boundaries and is innocuous since $P(U\in\{0,1\})=0$. Remark also that the approach proposed in this paper would still work if the quantile of $\log T(z)$ were linear only in a neighbourhood of the quantile $u$ of interest. We only assume linearity for all quantiles in $(0,1)$ in order to simplify the exposition. 

The random vector $Z$ can contain both exogenous and endogenous variables. We possess an instrumental variable, denoted by $W$, with support $\mathcal{W}\subset \mathbb{R}^L$, that is independent of $U$. All the exogenous variables in $Z$ are included in $W$. The distributions of $Z$ and $W$ can be discrete or continuous. We also assume that the distribution of $U$ given $Z$ and $W$ is absolutely continuous. Note that, by the inverse function theorem and equation \eqref{eq:model}, this continuity condition guarantees that the distribution of $T$ given $Z$ and $W$ is absolutely continuous too.

 In addition, the outcome variable $T$ is considered to be right censored by a random variable $C$ with values in $\mathbb{R}_{+}$. Define $Y = \min(T,C)$ and $\delta = \mathds{1}\{ T\le C\}$, where $\mathds{1}\{\cdot\}$ denotes the indicator function. The observables are $(Y,\delta, Z,W)$. 

Note that our model \eqref{eq:model} implies that 
\begin{equation}\label{eq11}\log T(z) = z^\top\beta_0(u) + \epsilon(u), \end{equation}
where $P(\epsilon(u)\le 0|W) = u$. Indeed, defining $\epsilon(u)= \log T(z) - z^\top \beta_0(u)$, we have $P(\epsilon(u)\le 0|W)= P(z^\top\beta_0(U)\le z^\top\beta_0(u)|W) = P(U\le u|W)=u$. Formulation \eqref{eq11} corresponds to the way the quantile regression model is often written in the literature, see e.g. \citet{honore2002quantile,khan2009inference,wang2009locally}. 

We now summarize the conditions we imposed in this section, for the value $u\in(0,1)$ of interest. 
\begin{assumption}\label{conditionR}
	(a) The mapping $\beta_0(\cdot)$ is continuously differentiable everywhere with derivative denoted by $D\beta_0(\cdot)$ and for all $u\in (0,1)$ and $z \in \mathcal{Z}$, $z^\top D\beta_0(u)>0$; (b) The distribution of $U$ given $Z$ and $W$ is absolutely continuous with continuous density; (c) $W\independent U$, where $\independent$ means “is independent of".
\end{assumption}

\section{Identification} 
\setcounter{equation}{0}
\setcounter{theorem}{0}
Now, we consider the identification of the parameter of interest $\beta_0(u)$. Using equation \eqref{eq:model} and the condition that $W \independent U$, it is possible to show that $\beta_0(u)$ is the solution to a continuum of identification equations, presented in the next subsection. This argument will lead to the aimed identification results, for both the case of censored and uncensored outcomes.

\subsection{Identification Equations}
We use model \eqref{eq:model} to show that $\beta_0(u)$ is a solution of the following system of equations in $\beta \in\mathbb{R}^K$:
\begin{equation}
    \label{eq:estimation}
    E[\mathds{1}\{T\le\exp(Z^\top\beta)\}|W=w] = u,\ \text{for all } w\in\mathcal{W}.
\end{equation}
Indeed, using, the specification of $T$ in model \eqref{eq:model}, Assumption \ref{conditionR} (a), and the fact that $U|W\sim U[0,1]$, we obtain the following equalities:
\begin{align*}
    E[\mathds{1}\{T\le\exp(Z^\top\beta_0(u))\}|W=w]  &= E[\mathds{1}\{\exp(Z^\top\beta_0(U))\le\exp(Z^\top\beta_0(u))\}|W=w] \\
        &=E[\mathds{1} \{ U \le u\}|W=w]\\
        &= u.
\end{align*}
{\ifcorr\color{red} \else \color{black} \fi  In the next sections, we derive identification results for the parameter of interest based on the system of equations \eqref{eq:estimation}}. 

\subsection{Identification without Censoring}
In this section, to simplify the exposition, we derive identification results without taking into account the censoring mechanism ($C=\infty$). In this case, the identification properties are more readily obtained since the left-hand side of Equation \eqref{eq:estimation} is identified. This implies that, in this context, studying identification is equivalent to assessing the uniqueness of the solutions to \eqref{eq:estimation}. Identification with censoring is discussed in Section \ref{sec.idcens}. 

The present section contains two types of identification results (both without censoring as mentioned in the above paragraph). We first show identification under a general framework in Section \ref{sec:globaId}. The identification conditions in Section \ref{sec:globaId} are technical and abstract, but standard in the literature on instrumental variable quantile regression models (see \citet{chernozhukov2005iv} or \citet{feve2018estimation}, among others). Next, we provide simpler and more interpretable identification results in the case of randomized experiments with noncompliance in Section \ref{subsec.nonc}.

\subsubsection{General identification result} \label{sec:globaId}
Let the parameter space $S$ be the set of continuously differentiable mappings $\beta(\cdot):(0,1)\xrightarrow{}\mathbb{R}^K$ with derivative denoted by $D\beta(\cdot)$ such that for all $u\in (0,1)$ and $z \in \mathcal{Z}$, $z^\top D\beta_0(u)>0$. The parameter $\beta_0(u)$ is identified (when there is no censoring) if, for $\beta(\cdot)\in S$,  $$
E[\mathds{1}\{T\le\exp(Z^\top\beta(u))\}|W=w] = u\quad \text{for all}\quad w\in\mathcal{W},
$$ implies that $\beta(u)=\beta_0(u)$. 

Rewrite the conditional distribution function $R(u|z,w)$ and density function $r(u|z,w)$ of $U$ given $Z=z,W=w$ as
\begin{align*}
    R(u|z,w) &= F_{T|Z,W}(\exp(z^\top\beta_0(u))|z,w);\\
    r(u|z,w) &= z^\top D\beta_0(u)\exp(z^\top\beta_0(u))f_{T|Z,W}(\exp(z^\top\beta_0(u))|z,w),
\end{align*}
where, $F_{T|Z,W}$ and $f_{T|Z,W}$ are, respectively, the cumulative distribution function and density of $T$ given $Z,W$. 

Let $\Delta$ be a mapping from $\mathcal{Z}\times(0,1)$ to $\mathbb{R}$ differentiable in its second argument. For $z\in\mathcal{Z}$ and $u\in (0,1)$, we use the notation $\Delta_z(u)=\Delta(z,u)$. The derivative of $\Delta_z(\cdot)$ with respect to $u$ is denoted by $D\Delta_z(\cdot)$. For $\alpha\in[0,1]$, define the $\alpha$-perturbation of the quantities $R(u|z,w)$ and $r(u|z,w)$ in the direction $\Delta_z(u)\in \mathbb{R}$ as 
\begin{align*}
    \tilde R_{\alpha,\Delta}(u|z,w) &= F_{T|Z,W}(\exp(z^\top\beta_0(u)) + \alpha^\top \Delta_z(u)|z,w);\\
    \tilde r_{\alpha,\Delta}(u|z,w)  &= [z^\top D\beta_0(u)\exp(z^\top\beta_0(u)) + \alpha^\top D\Delta_z(u)] f_{T|Z,W}(\exp(z^\top\beta_0(u)) + \alpha^\top \Delta_z(u)|z,w).
\end{align*}

We now assume that $Z$ is strongly complete by $W$ given $U=u$. 

\begin{definition} \label{def:strongcomplete}
The variable $Z$ is said to be strongly complete by $W$ given $U=u$, if for all families $\{\rho_{\alpha,\Delta}\}$ of functions from $\mathcal{Z}$ to $\mathbb{R}$ indexed by $\alpha$ and $\Delta$, the fact that
$$
\int_0^1 E\left[\left.\rho_{\alpha,\Delta}(Z)\tilde r_{\alpha,\Delta}(u|Z,w)\right|W=w\right]d\alpha = 0,
$$
for all $w\in\mathcal{W}$ and $\Delta_z$ implies that  $\rho_{\alpha,\Delta}\equiv 0$.
\end{definition}
This type of strong completeness condition is studied in \citet{chernozhukov2005iv}, both in the case where $Z$ and $W$ are continuous and the case where they are both categorical. It is an abstract condition requiring a certain degree of dependence between $Z$ and $W$. The next theorem contains the global identification result. 
\begin{theorem} \label{theo:globalIndentificationUncensored}
If $Z$ is strongly complete by $W$ given $U=u$, if Assumption \ref{conditionR} holds and if $E[ZZ^\top]$ is of full rank, then $\beta_0(u)$ is identified.
\end{theorem}
By identification here, we mean that $\beta_0(\cdot)$ is the unique solution in $S$ of the system of equations \eqref{eq:estimation}. Note that we require that $E[ZZ^\top]$ has full rank. {\ifcorr\color{red} \else \color{black} \fi We need this assumption due to the linear relation between $\log(T)$ and the covariate $Z$ in model \eqref{eq:model}.}

\subsubsection{Simpler identification conditions in randomized experiments with noncompliance}\label{subsec.nonc}
The conditions of Theorem \ref{theo:globalIndentificationUncensored} are not minimal. Indeed, in some special cases of applied relevance, it is possible to obtain simpler and more interpretable identification conditions by following arguments similar to that of \citet{chernozhukov2005iv}.
We focus on the case where the treatment $Z$ and the instrument $W$ are binary. Such a setting corresponds to randomized experiments with noncompliance which naturally occur in empirical applications. An example is the JTPA experiment discussed in Section \ref{sec:emp2}. Note that, it would be possible to include an intercept or exogenous covariates in the analysis but, for simplicity, we decided to avoid it.


Let $\nu,\underline{f}>0$ be small constants. We define the set $\mathcal{L}$ by the closed rectangle of vectors $(t_0,t_1)\in\mathbb{R}^2$ satisfying the following conditions, where $t_{Z}=(1-Z)t_0+ Zt_1$:
\begin{itemize}
        \item[(i)] $P(T\le\exp( t_{Z})|W=w_1)\in[u-\nu, u+\nu]$ for each $w_1\in\{0,1\}$,
        \item[(ii)] $f_{T|Z,W}(\exp(t_Z)|z,w)>\underline{f}$ for all $z,w\in\{0,1\}$ such that $P(Z=z|W=w)>0$.
    \end{itemize}

We want to show that there exists a unique $t=(t_0,t_1)^\top \in \mathcal{L}$ such that $\Pi(t)=0$, where 
$$
\Pi(t) = \begin{bmatrix}
P\big(T\le \exp(t_{Z})\big|W=0\big)-u\\
P\big(T\le \exp(t_{Z})\big|W=1\big)-u
\end{bmatrix}.
$$
The Jacobian of $\Pi(t)$ with respect to $t$, denoted by $\Pi'(t)$, takes the following form:
\begin{align*}
 \Pi'(t)= \begin{bmatrix}
f_{T,Z|W}(\exp(t_0),0|0)\exp(t_0) & f_{T,Z|W}(\exp(t_1),1|0)\exp(t_1)\\
f_{T,Z|W}(\exp(t_0),0|1)\exp(t_0) & f_{T,Z|W}(\exp(t_1),1|1)\exp(t_1)
\end{bmatrix},
\end{align*}
where $f_{T,Z|W}(t,z|W)= f_{T|Z,W}(t|z,w)P(Z=z|W=w).$ We make the following Assumption.
\begin{assumption} \label{conditionB}
	(a) $f_{T|Z,W}(\exp(z\beta_0(u))|z, w)>\underline{f}$ for all $z, w\in\{0,1\}$ such that $P(Z=z|W=w)>0$; (b) $\Pi'(t)\text{ is continuous in $t$ and has full rank for all }t\in\mathcal{L}$.
\end{assumption}

Assumption \ref{conditionB} (a) guarantees that $(0,\beta_0(u) )$ belongs to $\mathcal{L}$. It means that for every $z,w$ the density of $U$ given $Z=z,W=w$ is strictly positive. 
Assumption \ref{conditionB} (b) is a full rank condition, which is standard in econometrics. Similarly as in \citet{chernozhukov2005iv}, we can provide the following interpretation of Assumption \ref{conditionB} (b). The matrix $\Pi'(t)$ has full rank for all $t\in\mathcal{L}$ if and only if
\begin{equation} \label{slopes2new}
    \frac{f_{T,Z|W}(\exp(t_1),1|0)}{f_{T,Z|W}(\exp(t_0),0|0)} > \frac{f_{T,Z|W}(\exp(t_1),1|1)}{f_{T,Z|W}(\exp(t_0),0|1)},\text{ for all $t\in\mathcal{L}$}
\end{equation}
(or the same property with $<$ instead of $>$). This inequality can be interpreted as a monotone likelihood ratio condition:  for all $t\in\mathcal{L}$, the instrument increases (or decreases) the likelihood ratio in \eqref{slopes2new}.

Let us also consider the special case of randomized experiments with one-sided noncompliance where $P(Z = 1 | W = 0) = 0$ (this equality approximately holds in the JTPA empirical application). Then, it can be seen that Assumption \ref{conditionB} holds as long as $$P(Z = 1 | W = 1, U = u) > 0.$$
This condition means that subjects for which $U=u$ have a strictly positive probability to be treated when assigned to the treatment group. 

We have the following theorem.
\begin{theorem}\label{theorem:binary} Under Assumptions \ref{conditionR} and \ref{conditionB}, $\beta_0(u)$ is identified. 
\end{theorem}
Notice that a similar result would hold in the case where there are additional exogenous covariates $X$, that is when $Z = (Z_1,X^\top)^\top$, where $Z_1\in\{0,1\}$ and $W = (W_1,X^\top)^\top$, where $W_1\in\{0,1\}$. The only difference would be that all the probabilities and densities in the definition of $\mathcal{L}$ and Assumption \ref{conditionB} would be conditional on $X=x$ (for $x$ in the support of $X$), Assumption \ref{conditionB} would need to hold for all $x$ in the support of $X$, and we would have to assume also that $E[XX^\top]$ has full rank.

\subsection{Identification with Censoring}\label{sec.idcens}
In this section, we extend the identification results previously presented to the case where there is censoring. We assume that the censoring time $C$ satisfies the following assumption.
\begin{assumption}\label{conditionC}
	The censoring variable $C$ is independent of $(Z,W,U)$.  
\end{assumption}
Assumption \ref{conditionC} in particular implies that $T$ and $C$ are independent. {\ifcorr\color{red} \else \color{black} \fi We could replace it }by $C\independent U|Z,W$ but the latter assumption would complexify estimation and identification. Denote by $\bar c$ the upper bound of the support of $C$, and define $\bar u$ as 
\begin{align}
\label{eq:baru}
 \bar u = \inf\{u\in(0,1): \exists z\in \mathcal{Z} \text{ such that } \exp(z^\top\beta_0(u))> \bar c\}.
\end{align}
As we now explain, if $\bar u>0$, it is possible to identify the left-hand side of equation \eqref{eq:estimation} for any value of $u\in(0,\bar u)$. Let $\beta\in \mathbb{R}^K$ be such that $\exp(z^\top \beta)< \bar c$ for all $z\in\mathcal{Z}$. We have
\begin{align}\label{eq:equivalence}
E\Bigg[\frac{\delta}{G(Y)}
\mathds{1} \{Y\le \exp(Z^\top\beta)\}\bigg|W=w\Bigg] =     E[\mathds{1}\{T\le\exp(Z^\top\beta)\}|W=w],
\end{align}
where $G(s)=P(C\ge s)$ is the survival function of the censoring variable $C$. Note that, by definition of $\bar c$, we have $G(Y)>0$ on the event $\{Y\le \exp(Z^\top\beta)\}$. Hence, the left-hand side of \eqref{eq:equivalence} is well-defined.

We can justify equation \eqref{eq:equivalence}, using that $\delta = 1$ corresponds to $Y = T$ and so
{\allowdisplaybreaks \begin{align*}
    &E\Bigg[\frac{\delta}{G(Y)}
\mathds{1} \{Y\le \exp(Z^\top\beta)\}\Bigg|W=w\Bigg] \\
    &= E\Bigg[\frac{\mathds{1}\{T\le C\}\mathds{1}\{T\le \exp(Z^\top\beta)\}}{G(T)} \Bigg|W=w\Bigg] \\
    &= E\Bigg[\frac{\mathds{1}\{T\le \exp(Z^\top\beta)\}}{G(T)} E\bigg[\mathds{1}\{T\le C\} \bigg|T,Z,W\bigg]\Bigg|W=w\Bigg] \\
    &=E\left[\left.\mathds{1}\{T\le\exp(Z^\top\beta)\}\right|W=w\right],
\end{align*}}

\noindent  where in the last equality we use Assumption \ref{conditionC}, which implies that \linebreak $E[\mathds{1}\{T\le C\} |T,Z,W] = G(T)$. {\ifcorr\color{red} \else \color{black} \fi We can give the following interpretation of equality \eqref{eq:equivalence}}. The right-hand side of \eqref{eq:equivalence} is an average of indicator functions $\mathds{1}\{T\le\exp(Z^\top\beta)\}$ which are only observed when $\delta=1$. We can replace this average of $\mathds{1}\{T\le\exp(Z^\top\beta)\}$ by the average of the observed indicators $ \mathds{1}\{Y\le\exp(Z^\top\beta)\}$, weighted by $\delta/G(Y)$ to make them representative of the full sample.

Notice that $G(t)$ is identified (from the distribution of $(Y,\delta)$) for all $t\in[0,\bar c]$ by standard arguments from the survival analysis literature.
Hence, since $(Y,\delta, Z,W)$ are observed, equation \eqref{eq:equivalence} implies that the left-hand side of equation \eqref{eq:estimation} is identified for all $\beta\in \mathbb{R}^K$ such that $\exp(z^\top \beta)< \bar c$ for all $z\in\mathcal{Z}$, and so, in particular, for $\beta_0(u)$ for all $u\in (0,\bar u)$. Therefore, all the identification results previously discussed can be readily adapted to obtain identification under censoring for all $u\in (0,\bar u)$.

\section{Estimation Procedure}\label{sec:est}
\setcounter{equation}{0}
\setcounter{theorem}{0}
In this section, we provide an estimation procedure for the parameter vector $\beta_0(u)$. We possess an i.i.d. sample of size $n$ of the observables $(Y,\delta,Z, W)$, denoted by $\{(Y_i,\delta_i, Z_i, W_i)\}_{i=1}^n$.\\ 
Under Assumption \ref{conditionC} and using equation \eqref{eq:equivalence}, equation \eqref{eq:estimation} is equivalent to 
$$ E\Bigg[\frac{\delta}{G(Y)}
\mathds{1} \{Y\le \exp(Z^\top\beta(u))\}\bigg|W=w\Bigg] = u,\ \text{for all } w\in\mathcal{W}.$$
This continuum of conditional moment restrictions is equivalent to the following system of unconditional moment restrictions:
\begin{equation}\label{eq:estimation2}
E\Bigg[\frac{\delta}{G(Y)}
\mathds{1} \{Y\le \exp(Z^\top\beta(u)), W\le w\}\Bigg] =  uP(W\le w),\ \text{for all } w\in\mathcal{W},
\end{equation}
where by the event $\{W\le w\}$, we mean $\cap_{\ell=1}^L  \{W_\ell\le w_\ell\}$.
This new set of equations allows to design an estimation procedure which does not involve complex smoothing techniques. Let us define the following operator:
$$
\mathcal{A}_u(\beta,w) = E\Bigg[\frac{\delta}{G(Y)}
\mathds{1} \{Y\le \exp(Z^\top\beta), W\le w\}\Bigg] - uP(W\le w).
$$    
Consider now the following estimator $\hat{\mathcal{A}}_u(\beta,w)$ of $\mathcal{A}_u(\beta,w)$:
$$
\hat{\mathcal{A}}_u(\beta,w) =
\frac{1}{n}\sum_{i=1}^n \frac{\delta_i}{\hat G(Y_i)} \mathds{1} \{Y_i\le\exp(Z^\top_i\beta) , W_i\le w\} -  \frac{u}{n}\sum_{i=1}^n\mathds{1}(W_i\le w),
$$
where $\hat G(t)$ is the Kaplan-Meier estimator of $G$, that is
\begin{align*}
    \hat G(t) = \prod_{s\le t} \bigg(1-\frac{dN(s)}{Y(s)}\bigg),
\end{align*}
with $N(t) = \sum_{i=1}^n\mathds{1}\{Y_i\le t, \delta_i = 0\}$, $dN(s) = N(s)-\lim_{s'\xrightarrow{}s,s'<s}N(s')$ which denotes the jump of the process $N$ at the time $s$, and $Y(s) = \sum_{i=1}^n\mathds{1}(Y_i\ge s)$. Note that $\hat{\mathcal{A}}_u(\beta,w)$ is an average using only the uncensored observations which are made (approximately) representative of the full sample thanks to the weights $\{\delta_i/\hat{G}(Y_i)\}_{i=1}^n$. 
Our estimator $\hat\beta(u)$ of $\beta_0(u)$ is
\begin{equation}
\label{eq:estimamtorDefinition}
    \hat\beta(u) \in \argmin_\mathcal{\beta \in B} \frac{1}{n}\sum_{j=1}^n \hat{\mathcal{A}}^2_u(\beta,W_j), 
\end{equation}
where $\mathcal{B}$ is a compact parameter set in $\mathbb{R}^K$. This is a minimum distance estimator which attempts to solve estimated versions of the identification equation \eqref{eq:estimation2} at the points $W_1,\dots,W_n$. Minimum distance estimators have been studied in Econometrics, see for instance \citet{brown2002weighted, poirier2017efficient, torgovitsky2017minimum}. Our estimator has three distinctive features with respect to these papers. It concerns our specific semiparametric instrumental variable model, it deals with censoring and it solves the equation at the empirical distribution of $W$ rather than at a prespecified distribution. The last property avoids to let the econometrician choose the points (and the weights of these points) at which the equations are solved. {\ifcorr\color{red} \else \color{black} \fi We may be able to construct an alternative estimator attaining the semiparametric efficiency bound using the approach of \citet{poirier2017efficient}}. This estimator would however be more complicated and rely on features that are chosen by the econometrician.

Define the map $L:\mathcal{B}\xrightarrow{}\mathbb{R}$, where
$$
{L}(\beta) = E\left[ \mathcal{A}^2_u(\beta,W)\right],
$$
and let $\bar t$ be the upper bound of the support of $T$. Consider also the following condition.
\begin{assumption} \label{conditionE}
		(a) The parameter $\beta_0(u)$ is an interior point of $\mathcal{B}\subset \mathbb{R}^K$; (b)  The parameter space $\mathcal{B}$ is compact; (c) For all $u,w$, the map $\beta \xrightarrow{} \mathcal{A}_u(\beta,w)$ is three times differentiable at every point $\beta$ in the interior of $\mathcal{B}$ and all its third order derivatives are bounded uniformly in $u,w,\beta$; (d) The matrix $\nabla^2 L(\beta_0(u))$ is positive definite; (e) It holds that $\sup_{\beta\in \mathcal{B},z\in\mathcal{Z}} \exp(z^\top \beta)<\min(\bar c,\bar t)$; (f) The density of $T$ at $t$ given $Z = z$, namely $f_{T|Z}(t|z)$, is uniformly bounded in $(t,z)\in \mathbb{R}_+\times \mathcal{Z}$; (g) There exist a constant $\epsilon> 0$ and a function $b(\cdot):\mathcal{Z}\rightarrow{}\mathbb{R}$, such that $E[b^2(Z)]<\infty$ and $\|\exp(z^\top\beta)z^\top\|\le b(z)$ for all $z\in\mathcal{Z}$ and $\beta \in 	\{\beta\in \mathbb{R}^K: \|\beta - \beta_0(u )\|<\epsilon\}$.
\end{assumption}

In Assumption \ref{conditionE} (g), $\|\cdot\|$ corresponds to the Euclidean norm. Assumptions \ref{conditionE} (a), (b) are standard in the literature on $M-$estimators (see \citet{newey1994large}). Assumptions \ref{conditionE} (c), (d) and (f) are mild conditions which depend simultaneously on the regularity of the map $\beta \xrightarrow{} \exp(z^\top \beta)$ and on the distribution of $(U,Z,W)$. Note that  Assumption \ref{conditionE} (g) holds when $\mathcal{Z}$ is compact but is more general.  Assumption \ref{conditionE} (e) ensures that $\mathcal{A}_u(\beta,w)$ is identified at all $\beta\in\mathcal{B}$ (another role of this assumption is to avoid running into consistency problems of the Kaplan-Meier estimator on the tails). 
We have the following theorem.
\begin{theorem}
\label{theorem:2}
Under Assumptions \ref{conditionR}, \ref{conditionC}  and \ref{conditionE}, if $\beta_0(u)$ is identified, then there exists a $K\times K$ matrix $\Sigma$ such that
$$
\sqrt{n}(\hat{\beta}(u)-\beta_0(u)) \stackrel{d}{\rightarrow} \mathcal{N}(0,\Sigma).
$$
\end{theorem}
\subsection{Bootstrap} \label{sec:boot}
We present now a bootstrap procedure for inference regarding the proposed estimator. It avoids estimating its asymptotic variance matrix. Denote by $\{(Y_{bi},\delta_{bi}, Z_{bi}, W_{bi})\}_{i=1}^n$ a bootstrap sample drawn with replacement from the original sample $\{(Y_i,\delta_i, Z_i, W_i)\}_{i=1}^n$. In addition, denote by $\hat \beta_b(u)$ the value of the estimator computed on the bootstrap sample. The following asymptotic result holds.

\begin{theorem} \label{theo:boot}
Under Assumptions  \ref{conditionR}, \ref{conditionB}, and \ref{conditionC}, if $\beta_0(u)$ is globally identified, then 
$$
\sqrt{n}\{\hat{\beta}_b(u)-\hat{\beta}(u)\} \stackrel{d}{\rightarrow} \mathcal{N}(0,\Sigma) \quad\quad [P]
$$
where the convergence is for the law of $\hat\beta_b$ conditional on the original sample, in probability with respect to the original sample. Formally, 
$$
\sup_t \Bigg| P^*\big(\sqrt{n}\{\hat\beta_b(u)-\hat\beta(u)\}\le t\big) - P\big(\sqrt{n}\{\hat\beta(u)-\beta_0(u)\}\le t\big)\Bigg| =o_p(1), 
$$
where $P^*$ stands for the probability in law conditionally on the original data.
\end{theorem}

\section{Simulations}
\setcounter{equation}{0}
\setcounter{theorem}{0}
In this section, we present the results of numerical experiments to analyze the performance of the proposed estimator in finite samples. The data generating process (DGP) is as follows. A random variable $U\sim U[0,1]$ determines the structural quantile. Then,  $Z = (1,Z_2,Z_3)^\top$ is a $3\times 1$ random vector, where the component $Z_2$ of $Z$ is endogenous and the component $Z_3$ is exogenous i.e. $Z_3\independent U$ but  $Z_2\ \cancel{\independent}\ U$. Moreover, $W = (1,W_2,Z_3)^\top$ is the instrumental variable. We consider three different designs for the distributions of $W_2, Z_2, Z_3$. They are specified in Table \ref{table:design}.  In the first and third design, the endogenous component $Z_2$ has a discrete distribution, while in the second it is a continuous variable.
\begin{table}[H]
\caption{\label{table:design} Specification of the simulation design }
\begin{center}
\begin{tabular}{|c|c|c|c|}
\hline
Design & $W_2$          & $Z_2$                            & $Z_3$  \\ \hline
1      & Exp(1)         & $\mathds{1}\{ W_2+0.5U-1 > 0 \}$ & U(0,1) \\ \hline
2      & LogNormal(0,1) & $W_2+0.5U$ + 0.2U(0,1)           & Exp(1) \\ \hline
3      & B(0.5)         & $\mathds{1}\{ W_2+0.5U-1 > 0 \}$ & U(0,1) \\ \hline
\end{tabular}
\end{center}
\footnotesize
\renewcommand{\baselineskip}{11pt}
\textbf{Note:}
Distributions of  $W_2,Z_2$ and $Z_3$ for the designs 1, 2 and 3, where Exp($\lambda$), U(0,1), LogNormal($\mu$,$\sigma^2$) and B$(p)$ indicate, respectively,  the exponential distribution with parameter $\lambda$ (i.e. mean $1/\lambda$), the standard uniform distribution, the log-normal distribution with parameters $\mu$ and $\sigma^2$ and the Bernoulli distribution with parameter $p$.
\end{table}
The parameter of interest is $\beta_0(u) = (u,u,u)^\top$, and the duration $T$  follows the model \eqref{eq:model}, so $ T = \exp(Z^\top \beta_{0}(U))$. For each design, the random censoring time $C$ follows an exponential distribution with parameter $\lambda$ (i.e. its mean is $1/\lambda$). We consider $\lambda\in\{0.0068,0.176\}$ in the first design, $\lambda\in \{0.0173,0.065\}$ in the second design and $\lambda\in\{0.07, 0.175\}$ in the third design. These values of $\lambda$ ensure that there are 20\% or 40\% of censored observations. Therefore, the support of $C$ is $\mathbb{R}_{+}$ and $\bar u = 1$, where $\bar u$ is defined in \eqref{eq:baru}.

\begin{table}[H]
\caption{\label{table:sim} Simulation results}


\resizebox{\textwidth}{!}{
\begin{tabular}{|c|c|c|c|rrr|c|ccc|rrr|c|} 
	\cline{5-15}
	\multicolumn{1}{l}{} & \multicolumn{1}{l}{} & \multicolumn{1}{l}{}  & \multicolumn{1}{l|}{} & \multicolumn{7}{c|}{$\hat \beta$}                                                                                                                                                & \multicolumn{4}{c|}{$\hat \beta_{CQR}$}                                                                                \\ 
	\cline{5-15}
	\multicolumn{1}{c}{} & \multicolumn{1}{c}{} & \multicolumn{1}{c}{}  &                       & \multicolumn{3}{c|}{Bias}                     & \multicolumn{1}{r|}{RMSE} & \multicolumn{3}{c|}{95\% Cov. Prob.}                                                                 & \multicolumn{3}{c|}{Bias}                     & RMSE                             \\ 
	\hline
	Design               & u                    & n                     & Cens. \%              & $\beta_{0,1}$ & $\beta_{0,2}$ & $\beta_{0,3}$ & \multicolumn{1}{r|}{}     & \multicolumn{1}{r}{$\beta_{0,1}$} & \multicolumn{1}{r}{$\beta_{0,2}$} & \multicolumn{1}{r|}{$\beta_{0,3}$} & $\beta_{0,1}$ & $\beta_{0,2}$ & $\beta_{0,3}$ & \multicolumn{1}{r|}{}  \\ 
	\hline
	\multirow{12}{*}{1}  & \multirow{4}{*}{.3}  & \multirow{2}{*}{500}  & 20\%                  & .006          & .009          & -.010         & .180                      & .902                            & .900                            & .858                             & -.042         & .199          & -.042         & .265                          \\
	&                      &                       & 40\%                  & .006          & .007          & -.009         & .197                      & .874                            & .854                            & .808                             & -.036         & .192          & -.065         & .266                      \\ 
	\cline{3-15}
	&                      & \multirow{2}{*}{1000} & 20\%                  & -.001         & .011          & .007          & .136                      & .964                            & .930                            & .950                             & -.050         & .198          & -.027         & .235                          \\
	&                      &                       & 40\%                  & -.001         & .008          & .009          & .149                      & .954                            & .946                            & .912                             & -.044         & .191          & -.046         & .235                         \\ 
	\cline{2-15}
	& \multirow{4}{*}{.5}  & \multirow{2}{*}{500}  & 20\%                  & .005          & .009          & -.011         & .205                      & .950                            & .964                            & .904                             & -.061         & .223          & -.052         & .295                          \\
	&                      &                       & 40\%                  & .008          & .006          & -.017         & .239                      & .908                            & .930                            & .826                             & -.056         & .215          & -.080         & .304                           \\ 
	\cline{3-15}
	&                      & \multirow{2}{*}{1000} & 20\%                  & .000          & .008          & .002          & .159                      & .946                            & .928                            & .940                             & -.066         & .224          & -.041         & .270                          \\
	&                      &                       & 40\%                  & .001          & .003          & .004          & .179                      & .948                            & .932                            & .960                             & -.061         & .213          & -.065         & .270                           \\ 
	\cline{2-15}
	& \multirow{4}{*}{.7}  & \multirow{2}{*}{500}  & 20\%                  & .004          & .004          & -.011         & .212                      & .942                            & .954                            & .910                             & -.064         & .183          & -.043         & .267                         \\
	&                      &                       & 40\%                  & .011          & -.013         & -.016         & .255                      & .884                            & .930                            & .792                             & -.068         & .175          & -.052         & .285                            \\ 
	\cline{3-15}
	&                      & \multirow{2}{*}{1000} & 20\%                  & .000          & .001          & -.001         & .161                      & .968                            & .958                            & .930                             & -.067         & .180          & -.035         & .233                         \\
	&                      &                       & 40\%                  & .000          & -.002         & .003          & .199                      & .938                            & .946                            & .884                             & -.074         & .171          & -.035         & .242                            \\ 
	\hline
	\multirow{12}{*}{2}  & \multirow{4}{*}{.3}  & \multirow{2}{*}{500}  & 20\%                  & .018          & -.005         & -.010         & .205                      & .924                            & .922                            & .924                             & -.049         & .042          & -.041         & .178                          \\
	&                      &                       & 40\%                  & .021          & -.007         & -.011         & .215                      & .876                            & .890                            & .902                             & .008          & -.004         & -.095         & .183                          \\ 
	\cline{3-15}
	&                      & \multirow{2}{*}{1000} & 20\%                  & .010          & -.005         & .002          & .144                      & .932                            & .952                            & .950                             & -.038         & .035          & -.043         & .134                         \\
	&                      &                       & 40\%                  & .010          & -.006         & .003          & .155                      & .942                            & .940                            & .928                             & .015          & -.006         & -.092         & .145                         \\ 
	\cline{2-15}
	& \multirow{4}{*}{.5}  & \multirow{2}{*}{500}  & 20\%                  & .021          & -.007         & -.013         & .245                      & .918                            & .950                            & .942                             & .020          & -.001         & -.107         & .213                        \\
	&                      &                       & 40\%                  & .017          & -.005         & -.007         & .270                      & .872                            & .924                            & .936                             & .079          & -.060         & -.173         & .282                  \\
	\cline{3-15}
	&                      & \multirow{2}{*}{1000} & 20\%                  & .005          & .000          & .002          & .166                      & .938                            & .968                            & .946                             & .014          & .005          & -.110         & .170                        \\
	&                      &                       & 40\%                  & .006          & -.001         & .002          & .188                      & .928                            & .966                            & .946                             & .071          & -.045         & -.180         & .243                          \\ 
	\cline{2-15}
	& \multirow{4}{*}{.7}  & \multirow{2}{*}{500}  & 20\%                  & -.005         & .005          & -.002         & .241                      & .884                            & .910                            & .926                             & .101          & -.067         & -.148         & .282                         \\
	&                      &                       & 40\%                  & .000          & .007          & .005          & .291                      & .878                            & .864                            & .896                             & .152          & -.153         & -.183         & .390                        \\ 
	\cline{3-15}
	&                      & \multirow{2}{*}{1000} & 20\%                  & .001          & .001          & .002          & .176                      & .932                            & .958                            & .952                             & .075          & -.035         & -.157         & .228                      \\
	&                      &                       & 40\%                  & .009          & -.002         & .004          & .241                      & .876                            & .928                            & .910                             & .091          & -.091         & -.193         & .305                         \\ 
	\hline
	\multirow{12}{*}{3}  & \multirow{4}{*}{.3}  & \multirow{2}{*}{500}  & 20\%                  & .007          & -.001         & -.005         & .197                      & .914                            & .928                            & .872                             & .013          & -.004         & -.028         & .170                         \\
	&                      &                       & 40\%                  & .006          & .000          & .002          & .215                      & .916                            & .914                            & .800                             & .017          & -.004         & -.047         & .173                           \\ 
	\cline{3-15}
	&                      & \multirow{2}{*}{1000} & 20\%                  & .005          & -.002         & -.005         & .145                      & .952                            & .952                            & .926                             & .009          & -.005         & -.025         & .120                           \\
	&                      &                       & 40\%                  & .006          & .001          & -.006         & .158                      & .930                            & .940                            & .884                             & .013          & -.006         & -.046         & .130                       \\ 
	\cline{2-15}
	& \multirow{4}{*}{.5}  & \multirow{2}{*}{500}  & 20\%                  & .004          & .002          & -.004         & .223                      & .922                            & .948                            & .918                             & .012          & -.008         & -.038         & .187                           \\
	&                      &                       & 40\%                  & .005          & .003          & -.003         & .259                      & .908                            & .924                            & .880                             & .014          & -.014         & -.057         & .208                            \\ 
	\cline{3-15}
	&                      & \multirow{2}{*}{1000} & 20\%                  & .002          & -.003         & .001          & .165                      & .950                            & .938                            & .964                             & .009          & -.011         & -.032         & .142                      \\
	&                      &                       & 40\%                  & .003          & -.002         & .002          & .187                      & .958                            & .926                            & .930                             & .012          & -.016         & -.056         & .154                      \\ 
	\cline{2-15}
	& \multirow{4}{*}{.7}  & \multirow{2}{*}{500}  & 20\%                  & .006          & -.009         & .001          & .221                      & .948                            & .944                            & .944                             & .009          & -.017         & -.035         & .185                           \\
	&                      &                       & 40\%                  & .010          & -.002         & -.014         & .271                      & .922                            & .900                            & .904                             & .004          & -.027         & -.036         & .212                           \\ 
	\cline{3-15}
	&                      & \multirow{2}{*}{1000} & 20\%                  & .002          & -.004         & -.003         & .164                      & .932                            & .936                            & .934                             & .004          & -.016         & -.027         & .139                        \\
	&                      &                       & 40\%                  & .004          & -.003         & -.007         & .203                      & .946                            & .942                            & .924                             & -.003         & -.024         & -.030         & .158                          \\
	\hline
\end{tabular}
}

\footnotesize
\renewcommand{\baselineskip}{11pt}
\textbf{Note:} Results based on 500 simulations for the estimation of $\beta_0(u)$ for $u\in\{0.3,0.5,0.7\}$ for designs 1, 2 and 3, using the proposed estimator ($\hat\beta$) and the estimator of \cite{wang2009locally} ($\hat\beta_{CQR}$). The value of Design, u, $n$, Cens. \%, Bias, RMSE, and 95\% Cov. Prob. indicates, respectively, the design, the value of $u$, the sample size, the percentage of censoring, the bias estimation, the RMSE estimation, and the  95\% coverage  probabilities. 
\end{table}

We estimate $\beta_0(u)$ for quantiles $u\in\{0.3,0.5,0.7\}$ and sample sizes $n\in\{500,1000\}$. We use the algorithm of \citet{nelder1965simplex} for the minimization of the objective function in \eqref{eq:estimamtorDefinition}. {\ifcorr\color{red} \else \color{black} \fi We search for the minimum} in the compact set $\mathcal{B}=[0, 1]^3$, and the algorithm starts from a random point taken in $\mathcal{B}$ (the initial value follows a uniform distribution on $\mathcal{B}$). {\ifcorr\color{red} \else \color{black} \fi The optimization
algorithm starts at 100 random values}. Each of these starting values leads to a local
minimum. The final estimate $\hat{\beta}(u)$ corresponds to the local minimum yielding the lowest value of the objective function in \eqref{eq:estimamtorDefinition}. In Table \ref{table:sim}, we report, for each component $\beta_{0,k}$, $(k=1,2,3)$ of $\beta_0$, the bias of the estimator, its root-mean-squared error (RMSE), and the {\ifcorr\color{red} \else \color{black} \fi coverage of $95\%$ confidence intervals constructed by bootstrap percentiles. These results are based on 500 replications.} The RMSE is defined as the squared of the average over the simulations of the euclidean distance between the estimator and $\beta_0(u)$. {\ifcorr\color{red} \else \color{black} \fi We compute the coverage} using the method proposed in \citet{giacomini2013warp} to speed up the simulations. Lastly, we display the bias and RMSE of an estimator for the standard censored quantile regression ($\hat\beta_{CQR}$) for which the method of \citet{wang2009locally} is used. This CQR estimator ignores the endogeneity issue. 

  We see from the results that the bias of the proposed estimator is close to zero. Moreover, when the sample size increases, the bias and the coverage probability converge to their respective theoretical values of 0 and 0.95. The results improve when the proportion of censored observations is lower, and the RMSE tends to increase for higher quantiles. 
We also observe that the performance of the estimator is similar across the different designs. As expected, the standard censored quantile regression estimator of the endogenous regressor is biased.
  
  \section{Application to the National JTPA Study}\label{sec:emp2}

In this section, we apply the proposed method to estimate the effect of publicly subsidized job training programs on unemployment durations. The data (\cite{jtpaDS}) is collected from a large-scale randomized experiment known as the National JTPA Study, designed to evaluate programs funded by the Job Training Partnership Act of 1982. Different authors have analysed this dataset, for instance \citet{abadie2002instrumental} and \citet{frandsen2015treatment}. 

The experiment started in 1987 and involved around 21,000 economically disadvantaged individuals, who were randomly assigned to a treatment group or a control group. The treated subjects were allowed to enrol in a JTPA-funded training program, while the individuals from the control group were not allowed to enrol for 18 months. Our application focuses on a subset of 802 individuals 
 consisting of non-white single mothers unemployed at the time of randomization, who were surveyed in a follow-up interview taking place between 1 and 3 years following the random assignment.

About 75\% of the subjects in the sample were assigned to the treatment group (524 subjects), and 35\% to the control group (278 subjects). Among the subjects assigned to the treatment group, about 65\% (339 out of 524) participated in a JTPA-funded program. Among the ones assigned to the control group, only 36 subjects (around 12\%) enrolled in a JTPA-funded program (they enrolled more than 18 months after randomization, since they were forbidden from enrolling before).

The outcome variable $T$ is the duration (in days) between treatment assignment and finding employment. We are interested in the effect of two covariates on $T$. The regressors are $Z=(1,Z_2,Z_3)^\top$, where $Z_2$ is an indicator for participation in a JTPA-funded program  ($Z_2=1$ if the subject participates, $Z_2=0$ otherwise) and $Z_3$ is the age of the subject. We call $Z_2$ the treatment and treat it as endogenous. The covariate age is treated as exogenous. The instrumental variable is $W=(1,W_2,Z_3)^\top$, where $W_2$ is an indicator for treatment group assignment. The validity of $W_2$ as an instrument for $Z_2$ can be justified by the facts that (i) the assignment is random, (ii) being assigned to treatment or control should have no impact on unemployment durations other than through participation in a JTPA-funded program, and (iii) being assigned to the treatment group and enrolling to the JTPA program are dependent.

The outcome variable is only observed for individuals who found a job before the follow-up survey, and it is censored for the other individuals. Hence, we observe $Y=\min(T,C)$ and $\delta= \mathds{1}\{T\le C\}$, where $C$ is the duration between the randomization and the follow-up interview. This censoring time varies between individuals and is therefore random. The follow-up surveys are initiated by the officers in charge of data collection. They contact the subjects participating in the experiment according to a set of rules depending only on the date of entry in the experiment. Hence, the censoring time is primarily determined by the date of randomization. As a result, as long as this date is independent of subjects characteristics, censoring should be independent. \citet{frandsen2015treatment} provides additional reasons why censoring is likely to be independent.

 The average (over uncensored observations) unemployment duration of treatment group members is 20 days lower than that of their control group counterparts. The proportion of censored observations is around 32\%. The mode of the unemployment duration (when it is observed) is 26 days and the median is 361 days. A large number of observations of duration is around 600 days. This corresponds to the time around which many follow-up interviews were held. Beyond this point, almost all observations are censored.

To compute the estimator $\hat\beta(u)$ while avoiding local minima, we initialize the optimization algorithm at 1,000 random initialization points. We then select the estimate which minimizes the objective function among the resulting vectors. This procedure is applied for each $u \in \{0.1,0.2,..., 0.9\}$. {\ifcorr\color{red} \else \color{black} \fi We can use the results of the analysis to conduct an informal joint test of our assumptions. Indeed, recall that our identification conditions at $u$ include the fact that \begin{equation}\label{followdec} \text{$\exp(z^\top\beta(u))\le\bar c$, for all $z\in\mathcal{Z}$}\end{equation} (see Section \ref{sec.idcens}). If our identification and estimation conditions were satisfied at $u$, then $\widehat{\beta}(u)$ would converge to $\beta(u)$ and therefore, by \eqref{followdec}, we should expect to have \begin{equation}\label{followdec2} \text{$\exp(Z_i^\top\hat{\beta}(u))\le\bar c$, $i\in\{1,\dots,n\}$},\end{equation} with high probability. Since \eqref{followdec2} does not hold for the quantiles $u\in\{0.7, 0.8, 0.9\}$, we conclude that our conditions are not satisfied for these quantiles.
In order to construct confidence intervals, we use the bootstrap approach discussed in Section \ref{sec:boot}. }

\begin{figure}[h]
  \includegraphics[width=.49\linewidth]{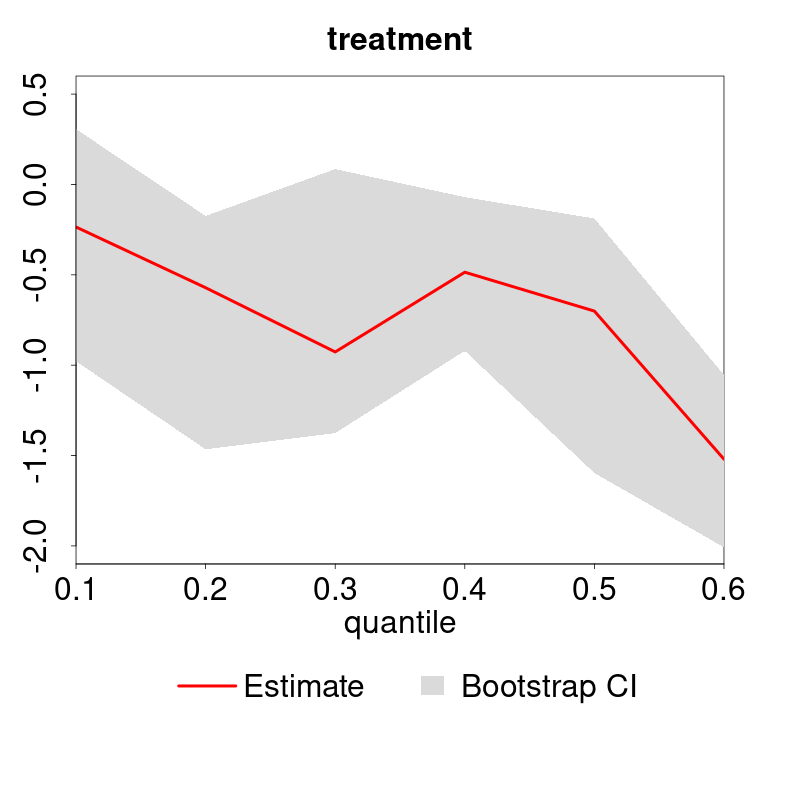}
  \includegraphics[width=.49\linewidth]{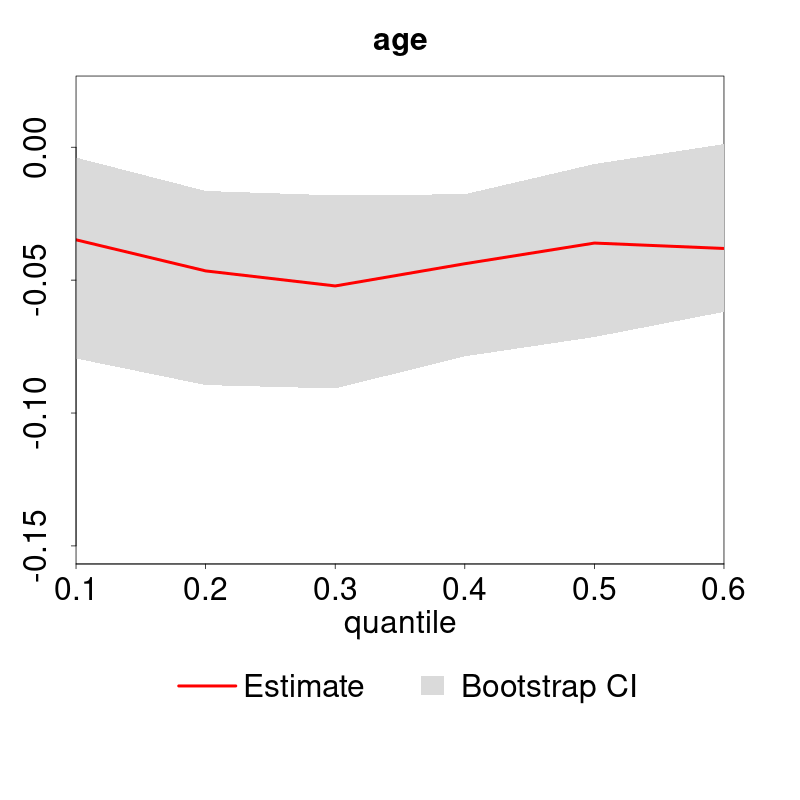}
\caption{\label{fig:figEmp2} Values of the proposed estimator $\hat\beta(u)$ and 95\% bootstrap confidence intervals.}
\label{fig:fig}
\end{figure}

{\ifcorr\color{red} \else \color{black} \fi We report the estimates and the 95\% (percentile bootstrap) confidence intervals for $\beta_2$ and $\beta_3$ in Figure \ref{fig:fig}.}
 The results indicate that the treatment has a significant and negative effect on $T$.  
 In addition, there is some evidence that the treatment effect varies by quantile. We find a significant and negative effect of age on unemployment duration. 
  In Section S3 of the supplementary material, we include a table reporting the exact values of the estimates and the bounds of the confidence intervals.
     \section*{Acknowledgements}
Financial support from the European Research Council (2016-2022, Horizon 2020 / ERC grant agreement No.\ 694409) is gratefully acknowledged.
\bibliographystyle{chicago}
\bibliography{paper-final}
\appendix
\section{Proof of identification results}

\subsection{Proof of Theorem \ref{theo:globalIndentificationUncensored}}
By \eqref{eq:estimation}, $\beta_0(\cdot)$  is a solution of 
$$ E\left[\left.F_{T|Z,W}(\exp(Z^\top\beta_0(u)) | Z, w)\right|W=w\right] = u,\ w\in\mathcal{W}.$$
Now suppose $\beta(\cdot)\in S$ is another solution. We obtain
$$E\left[\left.F_{T|Z,W}(\exp(Z^\top\beta_0(u)) | Z, w)-F_{T|Z,W}(\exp(Z^\top\beta(u)) | Z, w)\right|W=w\right]  = 0,\ w\in\mathcal{W}.$$
Denote by $\Delta_z(u)$ the quantity $\Delta_z(u) =\exp(z^\top\beta(u)) - \exp(z^\top\beta_0(u)).$ Then, the last equality can be rewritten as 
$$
E\left[\left.\int_0^1 f_{T|Z,W}(\exp(Z^\top\beta_0(u))+\alpha^\top \Delta_Z(u)|Z,w)\Delta_Z(u)d\alpha \right|W=w\right] = 0.
$$
Now, by definition of $S$, we have $z^\top D\beta(u)\exp(z^\top\beta(u)) >0$ for all $\beta\in S$ and $z\in\mathcal{Z}$.  This implies that for all $\alpha\in[0,1]$, it holds that $z^\top D\beta_0(u)\exp(z^\top\beta_0(u)) + \alpha^\top D\Delta_z(u)>0$,
where $D\Delta_z(u)= z^{\top}D\beta(u)\exp(z^\top\beta(u))-z^{\top}D\beta_0(u)\exp(z^\top\beta_0(u))$ is the differential of $\Delta_z(\cdot)$ evaluated at $u$. Therefore, using Fubini's Theorem and the definition of $\tilde r_{\alpha,\Delta}$ given in Section \ref{sec:globaId}, the last equation takes also the form
$$
\int_0^1E\left[\left. \frac{\Delta_Z(u)}{Z^\top D\beta_0(u)\exp(Z^\top\beta_0(u)) + \alpha^\top D\Delta_Z(u)}\tilde r_{\alpha,\Delta}(u|Z,w)\right|W=w\right]d\alpha = 0,
$$
The strong completeness of $Z$ by $W$ given $U=u$ implies that, almost everywhere,
$$\frac{\Delta_Z(u)}{Z^\top D\beta_0(u)\exp(Z^\top\beta_0(u)) + \alpha^\top D\Delta_Z(u)} = 0,$$ and so $\Delta_Z(u) = 0$ and $Z^\top \beta(u) = Z^\top \beta_0(u)$ almost surely. Taking the expected value on both sides we obtain $\beta(u) =\beta_0(u)$, since $\mathbb{E}[ZZ^\top]$ is full rank by hypothesis.

\subsection{Proof of Theorem \ref{theorem:binary}}

The proof is inspired by that of Theorem 2 in \citet{chernozhukov2005iv}. Consider the two iso-probability curves $\mathcal{T}$ and $\widetilde{\mathcal{T}}$, defined as 
\begin{align*}
    \mathcal{T} &= \big\{t = (t_0,t_1)\big| P(T\le \exp(t_{Z})|W=0)
  =u \big\}, \\
    \widetilde{\mathcal{T}} &= \big\{t = (t_0,t_1)\big| P(T\le \exp(t_{Z})|W=1) =u \big\}.
\end{align*}
Under Assumption \ref{conditionR}, the curves  $\mathcal{T}$ and $\widetilde{\mathcal{T}}$ satisfy the differential equations
\begin{align*}
     & f_{T,Z|W}(\exp(t_0),0|0)\exp(t_0)\text{d}t_0 + f_{T,Z|W}(\exp(t_1),1|0)\exp(t_1)\text{d}t_1  =0; \quad \text{and}\\
     & f_{T,Z|W}(\exp(t_0),0|1)\exp(t_0)\text{d}t_0 + f_{T,Z|W}(\exp(t_1),1|1)\exp(t_1)\text{d}t_1  =0.
\end{align*}
Therefore, the slopes at a given point point $(t_0,t_1)\in\mathcal{L}$ of the curves  $\mathcal{T}$ and $\widetilde{\mathcal{T}}$ are, respectively,
\begin{align}\label{slopes}
      \notag&\Big(\frac{\text{d}t_0}{\text{d}t_1}\Big)\Big\vert_{(t_0,t_1)} = -\frac{f_{T,Z|W}( \exp(t_1),1|0)\exp(t_1) }{  f_{T,Z|W}( \exp(t_0),0|0)\exp(t_0)}; \\& \widetilde{\Big(\frac{\text{d}t_0}{\text{d}t_1}\Big)}\Big\vert_{(t_0,t_1)} = -\frac{f_{T,Z|W}( \exp(t_1),1|1)\exp(t_1) }{  f_{T,Z|W}( \exp(t_0),0|1)\exp(t_0)}.
\end{align}
Since $\Pi'(t)$ has full rank and only positive entries, the slopes of the two curves evaluated at the same point are different from each other and non-positive. We now show that such condition is sufficient to obtain that  $\mathcal{T}\cap\widetilde{\mathcal{T}}\cap \mathcal{L}$ is a singleton. 
Because $\Pi'(t)$ is continuous in $t$, its determinant $\det \Pi'(t)$ is continuous too. The condition of full rank for $\Pi'(t)$ implies $\det \Pi'(t) > 0$ (or $<0$) for all $t \in \mathcal{L}$. This is equivalent to 
\begin{equation} \label{slopes2}
\begin{aligned}
    &\frac{f_{T,Z|W}(t_1,1|0)}{f_{T,Z|W}(t_0,0|0)} >  \frac{f_{T,Z|W}(t_1,0|1)}{f_{T,Z|W}(t_0,0|1)},
\end{aligned}
\end{equation}
or the same inequality with $<$ instead of $>$. Since $\mathcal{L}$ is compact, due to page 8 of \citet{milnor1997topology}, the set  $\mathcal{T}\cap\widetilde{\mathcal{T}}\cap \mathcal{L}$ is finite and equal to $\{t^{(1)},...,t^{(k)}\}$, with $k<\infty$ and  $t^{(i)} = (t_0^{(i)},t_1^{(i)})\in\mathbb{R}^2$. 

We rely on a proof by contradiction. Let us now assume that $k\ge2$. Since the slopes of the curves $\mathcal{T}$ and $\widetilde{\mathcal{T}}$ are not positive, crossing points of $\mathcal{T}$ and $\widetilde{\mathcal{T}}$ cannot be always from above or below. Hence, there exists at least two solutions $t^{(i)},t^{(j)}$ in $\mathcal{L}$ such that the slopes of the two curves at $t^{(i)},t^{(j)}$  satisfy the relations
\begin{align}\label{differentSlops}
      \Big(\frac{\text{d}t_0}{\text{d}t_1}\Big)\Big\vert_{t^{(i)}} >\widetilde{\Big(\frac{\text{d}t_0}{\text{d}t_1}\Big)}\Big\vert_{t^{(i)}} ,  \quad  \quad \text{and}       \quad  \quad        \Big(\frac{\text{d}t_0}{\text{d}t_1}\Big)\Big\vert_{t^{(j)}} <\widetilde{\Big(\frac{\text{d}t_0}{\text{d}t_1}\Big)}\Big\vert_{t^{(j)}}.
\end{align}
In other words, if there are multiple crossing points in $\mathcal{L}$ of the curves $\mathcal{T}$ and $\widetilde{\mathcal{T}}$, at least one of these crossing points is from above and at least another one is from below. However, \eqref{differentSlops} contradicts \eqref{slopes} and \eqref{slopes2}, so that $k=1$.

In conclusion, we obtain that $\mathcal{T}\cap\widetilde{\mathcal{T}}\cap \mathcal{L}$ is a singleton. This singleton has to correspond to $(0,\beta_0(u))$ since the latter belongs to $\mathcal{T}\cap\widetilde{\mathcal{T}}\cap \mathcal{L}$ by Assumption \ref{conditionB} (a). This implies that $(0,\beta_0(u))$ is identified because the set $\mathcal{T}\cap\widetilde{\mathcal{T}}\cap \mathcal{L}$ is identified from the distribution of the observables. \hfill $\Box$

\section{Proof of the results of Section 4 } 
\subsection{Proof of Theorem \ref{theorem:2}}
The main steps of the proof are presented in Section \ref{sec:main_proof}. They rely on technical lemmas from Section \ref{sec:lemmasas}. Since we consider a fixed value of $u$, without possible misunderstanding we write $\beta_0$ (resp. $\widehat{\beta}$) instead of $\beta_0(u)$ (resp. $\widehat{\beta}(u)$). In the proof we use the following quantities:
\begin{align*}
    \bar{{L}}(\beta) = n^{-1} \sum_{i=1}^n \mathcal{A}_u^2(\beta,W_i);\ \hat{{L}}(\beta) =n^{-1} \sum_{i=1}^n \hat{\mathcal{A}}^2_u(\beta,W_i).
\end{align*}
\subsubsection{Main proof} \label{sec:main_proof}
\textbf{Consistency} Under identification, $\beta_0$ is the unique minimizer of $L(\beta)$. Considering also Assumption \ref{conditionE} (b), which in particular implies that $L$ is continuous, and Lemma \ref{lemma:theo1lemma1} (c), we obtain that all the hypotheses of Theorem 2.1 of \citet{newey1994large} are satisfied, and so $\hat \beta - \beta_0= o_p(1)$.\\ 

\noindent \textbf{Asymptotic expansion} Consider a sequence of $\beta\in\mathcal{B}$ such that $\beta\xrightarrow{}\beta_0$. We have
\begin{align*}
    \hat{L}(\beta) - \bar{L}(\beta) &= n^{-1}\sum_{i=1}^n \left[ (\hat{\mathcal{A}}_u- \mathcal{A}_u+\mathcal{A}_u)(\beta,W_i)  \right]^2 - n^{-1}\sum_{i=1}^n \mathcal{A}^2_u(\beta,,W_i)\\
    &=n^{-1}\sum_{i=1}^n \left[ (\hat{\mathcal{A}}_u- \mathcal{A}_u)(\beta,W_i)  \right]^2 + 2n^{-1}\sum_{i=1}^n   \mathcal{A}_u(\beta,W_i)\left[ (\hat{\mathcal{A}}_u- \mathcal{A}_u)(\beta,W_i)\right].
\end{align*}
Thanks to Lemma \ref{lemma:theo1lemma1} (a) and (b) below, we obtain
\begin{align*}
    \hat{L}(\beta) - \bar{L}(\beta)=n^{-1}\sum_{i=1}^n \hat{\mathcal{A}}^2_u(\beta_0,W_i)+ 2n^{-1}\sum_{i=1}^n   \mathcal{A}_u(\beta,W_i)\left[ (\hat{\mathcal{A}}_u - \mathcal{A}_u)(\beta,W_i)\right]  + o_p(n^{-1}).
\end{align*}
This leads to 
\begin{align*}
    \hat{L}(\beta) - \bar{L}(\beta) 
     & =n^{-1}\sum_{i=1}^n \hat{\mathcal{A}}^2_u(\beta_0,W_i) +2(\beta-\beta_0)^\top n^{-1}\sum_{i=1}^n \frac{\partial \mathcal{A}_u}{\partial \beta}(\beta_0,W_i)\left[(\hat{\mathcal{A}}_u-\mathcal{A}_u)(\beta,W_i)\right]\\ 
    &\quad\quad + o_p(n^{-1}) + o_p(\|\beta-\beta_0\|^2)\\
    &= n^{-1}\sum_{i=1}^n \hat{\mathcal{A}}^2_u(\beta_0,W_i) +2(\beta-\beta_0)^\top n^{-1}\sum_{i=1}^n \frac{\partial \mathcal{A}_u}{\partial \beta}(\beta_0,W_i)\hat{\mathcal{A}}_u(\beta_0,W_i)\\
     &\quad\quad+ o_p(n^{-1}) + o_p(\|\beta-\beta_0\|^2) + o_p(\|\beta-\beta_0\|n^{-1/2}),
\end{align*}
where, in the first equality, we used a Taylor expansion and the uniform boundedness of the first two derivatives of the map $\beta \xrightarrow{}\mathcal{A}_u(\beta,w)$ at $\beta_0$ (Assumption \ref{conditionE} (c));  and, in the second equality, we apply again Lemma \ref{lemma:theo1lemma1}(a) and (b).
By definition of $\hat V $ in Lemma \ref{lemma:theo1lemma2}, we have
\begin{align*}
     \hat{L}(\beta) - \bar{L}(\beta) -   \hat{L}(\beta_0)  &= 2(\beta-\beta_0)^\top\hat V + o_p(n^{-1} + \|\beta-\beta_0\|^2 + \|\beta-\beta_0\|n^{-1/2}), 
\end{align*}
and, by the second order Taylor expansion of $\bar{L}(\beta)$ around $\beta_0$, it is also true that
\begin{align}
     \hat{L}(\beta) - \hat{L}(\beta_0)  &=2(\beta-\beta_0)^\top\hat V + (\beta-\beta_0)^\top \nabla ^2 \bar{L}(\beta_0)  (\beta-\beta_0) \nonumber\\
     &\quad + o_p(n^{-1} + \|\beta-\beta_0\|^2 + \|\beta-\beta_0\|n^{-1/2}).\label{eq:taylorExpansion}
\end{align}
Note that the $o_p$ in \eqref{eq:taylorExpansion} is uniform in the sequence of $\beta\in\mathcal{B}$ considered (this is because we only used arguments uniform in $\beta$). This uniformity is implicitly used in the remainder of the proof. We do not make this uniformity explicit to simplify the exposition.\\

\noindent \textbf{Rate of convergence}
By definition of $\hat \beta$, it holds that $\hat L(\hat \beta) - \hat L(\beta_0)\le 0$. Next, $\hat L(\hat \beta) - \hat L(\beta_0)\le 0$ and equation \eqref{eq:taylorExpansion} yield 
\begin{align}
\label{eq:defPos}
   &(\hat\beta-\beta_0)^\top \nabla ^2 \bar{L}(\beta_0)  (\hat\beta-\beta_0) \nonumber\\
    &\le |(\hat\beta-\beta_0)^\top2\hat V|  + o_p(n^{-1} + \|\hat\beta-\beta_0\|^2 + \|\hat\beta-\beta_0\|n^{-1/2}).
\end{align}
By Lemma \ref{lemma:theo1lemma2} below, we have $(\beta-\beta_0)^\top2\hat V = O_p(n^{-1/2}\|\beta - \beta_0\|)$. Moreover, by the law of large numbers, $\nabla^2\bar{L}(\beta_0)\xrightarrow{} \nabla^2L(\beta_0)$. Considering that   
$\nabla^2L(\beta_0)$ is positive definite by Assumption \ref{conditionE} (d), and the inequality in \eqref{eq:defPos}, it follows
$$
M\|\hat\beta - \beta_0\|^2 \le  O_p(n^{-1/2}\|\hat\beta - \beta_0\|) + o_p(n^{-1} + \|\hat\beta-\beta_0\|^2 + \|\hat\beta-\beta_0\|n^{-1/2}),
$$
where $M>0$ is a constant. This argument leads to
$$
\|\hat \beta - \beta_0\| = o_p(n^{-1/2}).
$$
\noindent \textbf{Asymptotic normality}
In order to conclude the proof, define $\mathcal{H}= \{h=\beta-\beta_0, \beta\in \mathcal{B}\}$. By \eqref{eq:taylorExpansion} and the fact that $\mathcal{B}$ is compact, we have
\begin{align*}
    n\left[ \hat{L}(\beta_0 + n^{-1/2}h) - \hat{L}(\beta_0) \right] =  2 n^{1/2} h^\top \hat{V} + h^\top  \nabla^2\bar{L}(\beta_0)h + R(h),
\end{align*}
where $\sup_{h\in\mathcal{H}}|R(h)| = o_p(1)$. Next, define $\hat U(h) = 2 n^{1/2} h^\top \hat{V} + h^\top \nabla^2\bar{L}(\beta_0)h $,  and  $\hat h_n = n^{1/2}(\hat\beta - \beta_0)$. Then, it holds that
$$
\inf_{h\in \mathcal{H}}  n\left[ \hat{L}(\beta_0 + n^{-1/2}h) - \hat{L}(\beta_0) \right] \le \inf \hat U(h) + o_p(1),
$$
This leads to
$$
\hat h_n \in \arg \min_{h \in \mathcal{H}}   n\left[ \hat{L}(\beta_0 + n^{-1/2}h) - \hat{L}(\beta_0) \right].
$$
Therefore, we have
$$
\hat U(\hat h_n) \le \inf_{h \in  \mathbb{R}^K } \hat U(h) +o_p(1).
$$
By the law of large numbers, $\nabla^2\bar L(\beta_0)\xrightarrow{}_P\nabla^2 L(\beta_0)$. Applying now Slutsky's lemma and Lemma \ref{lemma:theo1lemma2}, we obtain that $\hat U(\cdot) $ converges in distribution to $U(\cdot)$, with $U(h) = h^\top 2V + h^\top
\nabla^2L(\beta_0)h$. The score function $U(h)$ is uniquely maximised at $-\nabla^2L(\beta_0)^{-1}V$, which is a tight random variable. The sequence $\hat h_n$ is also uniformly tight since $\hat h_n = O_p(1)$.
Because of this, we can apply Theorem 14.1 in \citet{kosorok2008introduction} and conclude that $\hat h_n$ converges in distribution to $-\nabla^2L(\beta_0)^{-1}V$. 

\subsubsection{Technical lemmas}\label{sec:lemmasas}
\begin{lemma}
\label{lemma:theo1lemma1} Under Assumption \ref{conditionR}, \ref{conditionC}  and \ref{conditionE}, if $\beta_0(u)$ is identified, then:
\setcounter{bean}{0}
\begin{list}
	{(\alph{bean})}{\usecounter{bean}}
	\item   $\sup\limits_{ \beta \in \mathcal{B},w\in \mathcal{W}} \left| (\hat{\mathcal{A}}_u- \mathcal{A}_u)(\beta,w) \right| = O_p(n^{-1/2})$.
	\item $\lim\limits_{\epsilon \xrightarrow{}0  }\sup\limits_{\| \beta - \beta_0 \|\le \epsilon, w\in \mathcal{W}}  n^{1/2} \left| (\hat{\mathcal{A}}_u - \mathcal{A}_u)(\beta,w) - \hat{\mathcal{A}}_u(\beta_0,w)  \right| = o_p(1)$.
	\item $\sup\limits_{\beta \in \mathcal{B}} \left| \hat L(\beta) -L(\beta)\right| = o_p(1)$.
\end{list}
\end{lemma}

\textbf{Proof:} Let us first prove (a). Consider the operator $\tilde{\mathcal{A}}_u: \mathcal{B}\times \mathcal{W} \xrightarrow{}\mathbb{R}$, defined as
\begin{equation}\label{eqb4}
 \tilde{\mathcal{A}}_u(\beta,w) = \frac{1}{n} \sum_{i=1}^n \frac{\delta_i}{ G(Y_i)} \mathds{1} \{Y_i\le \exp(Z^\top_i\beta(u)) , W_i\le w\} -\frac{u}{n}\sum_{i=1}^n\mathds{1}(W_i\le w). 
\end{equation}
We have $(\hat{\mathcal{A}}_u - \mathcal{A}_u)(\beta,w) = \hat I_1(\beta,u,w) + \hat I_2(\beta,u,w),$ where $$\hat I_1(\beta,w)  = (\hat{\mathcal{A}}_u - \tilde{\mathcal{A}}_u)(\beta,w); \quad 
\hat{I}_2(\beta,w)  = (\tilde{\mathcal{A}}_u - \mathcal{A}_u)(\beta,w).$$

We now show that $$\sup\limits_{ \beta \in \mathcal{B},w\in \mathcal{W}} \left| \hat I_j(\beta,w) \right| = O_p(n^{-1/2}),\ j=1,2$$
to obtain the assertion. 

For $j=2$, notice that the class of functions
$$\mathcal{F}=\left\{\begin{array}{rl}\mathbb{R}\times\{0,1\}\times\mathcal{Z}\times\mathcal{W}&\mapsto  \mathbb{R}\\ (y,d,z,w)&\mapsto \frac{d}{G(y)}  \mathds{1}\{y \le \exp(z^\top\beta)\}-u\mathds{1}\{w\le \omega\} 
\end{array},\ \beta \in \mathcal{B},\ \omega\in\mathcal{W}\right\}$$
is Donsker. Indeed, it is the result of elementary operations of uniformly bounded Donsker classes (to see this use Lemma \ref{lemma:VCclass}). Therefore,
\begin{align*}
   \hat{I}_2(\beta,w) &= \frac{1}{n} \sum_{i=1}^n \frac{\delta_i}{G(Y_i)} \mathds{1} \{Y_i\le\exp(Z^\top_i\beta(u)) , W_i\le w\}\\&\quad -\mathbb{E}\left[\frac{\delta}{G(Y)}
\mathds{1} \{Y\le\exp(Z^\top\beta(u)), W\le w\}\right]\\
&\quad + \frac{u}{n}\sum_{i=1}^n\mathds{1}(W_i\le w) - uP(W\le w),
\end{align*}
is an empirical process over a family of Donsker functions and we get
$$\sup\limits_{ \beta \in \mathcal{B}, w\in \mathcal{W}} \left| \hat I_2(\beta,w) \right| = O_p(n^{-1/2}).$$\\
For $j=1$, consider the difference
\begin{align}\label{eq:KM1}
   \frac{1}{\hat G(t)} - \frac{1}{  G(t)} = \frac{\hat G(t)- G(t)}{G(t)\hat G(t)}.
\end{align}
Let $\tau = \inf\limits_{t\in \mathbb{R}}\{ \sup\limits_{z\in\mathcal{Z}, \beta \in \mathcal{B}}\exp(z^{\top}\beta)< t\} $. By Assumption \ref{conditionE} (e), we have that $\tau<\min(\bar c,\bar t)$. Hence, by standard properties of the Kaplan-Meier estimator $\hat G$, (see Theorem 1.1 in \citet{gill1983large}), it holds  \begin{align}\label{eq:KM2}
    \sup\limits_{t\in[0,\tau]}| G(t) - \hat G(t)| = O_p(n^{-1/2}),
\end{align}
From \eqref{eq:KM1} and \eqref{eq:KM2}, using that $G(\tau)>0$, it follows  that
 \begin{align*}
   \frac{1}{\hat G(Y_i)} - \frac{1}{ G(Y_i)} = O_p(n^{-1/2}).
\end{align*}
This leads to
\begin{align*}
   \hat{I}_1(\beta,u,w) &= \frac{1}{n} \sum_{i=1}^n \delta_i \mathds{1} \{Y_i\le\exp(Z^\top_i\beta) , W_i\le w\}\left(\frac{1}{\hat G(Y_i)} - \frac{1}{ G(Y_i)}\right)\\
   &= \frac{1}{n} \sum_{i=1}^n \delta_i \mathds{1} \{Y_i\le\exp(Z^\top_i\beta) , W_i\le w\}O_p(n^{-1/2})= O_p(n^{-1/2}).
\end{align*}

Now, we show (b). Consider again the quantities $\hat{I}_1(\beta,u,w)$ and $\hat{I}_2(\beta,u,w)$ previously defined. Let us prove that  
\begin{align*}
\lim\limits_{\epsilon \xrightarrow{}0  }\sup\limits_{\| \beta - \beta_0 \|\le \epsilon, w\in \mathcal{W}}   | \hat{I}_j(\beta,w)- \hat{I}_j(\beta_0,w)| = o_p(n^{-1/2}),\ \text{for } j=1,2,
\end{align*}
to obtain the assertion.\\
For $j=2$, we already noticed  that $\hat I_{2}(\beta,w)$ is an empirical process over a Donsker class. 
We now need to show that
\begin{align}\label{I2equi}
      \lim\limits_{\epsilon \xrightarrow{}0  }\sup\limits_{\| \beta - \beta_0 \|\le \epsilon, w\in \mathcal{W}}|\hat I_2(\beta,w) - \hat I_2(\beta_0,w)|&=o_p(1).
\end{align}
By the analysis in the proof of Lemma 3 of \citet{brown2002weighted}, \eqref{I2equi} follows from 
\begin{align}
    \label{eq:Kprop}
\lim\limits_{\epsilon \xrightarrow{}0  }\sup\limits_{\| \beta - \beta_0 \|\le \epsilon, w\in \mathcal{W}} \mathbb{E}\{[K(\beta,w) - K(\beta_0,w)]^2\} = 0,
\end{align}
where $K(\beta, w) =\frac{\delta}{G(Y)}\mathds{1} \{ Y\le \exp(Z^\top\beta ), W   \le w\} $. Therefore, it only remains to show the validity of equation \eqref{eq:Kprop}. We have
\begin{align*}
    \mathbb{E}\{[K(\beta,w)-K(\beta_0,w)]^2\}&=  \mathbb{E}\{K(\beta,w)[K(\beta,w)-K(\beta_0,w)]\}\\
    &\quad - \mathbb{E}\{K(\beta_0,w)[K(\beta,w)-K(\beta_0,w)]\}.
\end{align*}
Remark that $K(\beta,w)\le M$ where $M=1/G(\tau)$. This leads to
\begin{align*}
   & |\mathbb{E}\{K(\beta,w)[K(\beta,w)-K(\beta_0,w)]\}| \\
    &\le  M\mathbb{E}\{|K(\beta,w)-K(\beta_0,w)|\}\\
    &\le M\mathbb{E}\{|\mathds{1}\{ T\le \exp(Z^\top\beta)\} - \mathds{1}\{Y\le \exp(Z^\top\beta_0)\} |\}\\
    &=O(\|\beta -\beta_0\|), 
\end{align*}
where, in the last inequality, we used Lemma \ref{lemma:VCclass}. Hence, we obtain \eqref{eq:Kprop} and the assertion for $j=2$. For $j=1$, we can argue as follows:
\begin{align*}
    &|\hat{I}_1(\beta,w) - \hat{I}_1(\beta_0,w)|^2\\
    &\le    \sup_{w\in\mathcal{W},\beta\in\mathcal{B}} \frac{1}{n} \sum_{i=1}^n \delta_i| \mathds{1} \{Y_i\le\exp(Z^\top_i\beta)\} - \mathds{1} \{Y_i\le\exp(Z^\top_i\beta_0)\} |\\ 
    &\quad \times  \sup_{w\in\mathcal{W},\beta\in\mathcal{B}} \frac{1}{n} \sum_{i=1}^n \Bigg\{ \frac{\delta_i}{G(Y_i)\hat{G}(Y_i)}\bigg[\hat{G}(Y_i) - G(Y_i)\bigg]  \Bigg\}^2 \mathds{1} \{W_i\le w,Y_i\le\exp(Z^\top_i\beta)\vee \exp(Z^\top_i\beta_0)  \} \\
    &= \sup_{w\in\mathcal{W},\beta\in\mathcal{B}} \frac{1}{n} \sum_{i=1}^n | \mathds{1} \{T_i\le\exp(Z^\top_i\beta)\} - \mathds{1} \{T_i\le\exp(Z^\top_i\beta_0)\} | \times o_p(n^{-1}),
\end{align*}
where in the last equality we use again the asymptotic properties of the Kaplan-Meier estimator $\hat G$. 
Using Lemma \ref{lemma:VCclass}, we obtain
\begin{align*}
     \lim_{\epsilon \xrightarrow{}0} \sup_{ \|\beta -\beta_0\|\le \epsilon  }\frac{1}{n} \sum_{i=1}^n | \mathds{1} \{T_i\le\exp(Z^\top_i\beta)\} - \mathds{1} \{T_i\le\exp(Z^\top_i\beta_0)\} | = o_p(1).
\end{align*}
This, together with the previous inequality, leads to the result also for $j=1$.

Finally, we prove (c). It follows from (a) that $\sup_{\beta \in \mathcal{B}}|\hat L(\beta) - \bar L(\beta) | = o_p(1)$, and so it is enough to show that $\sup_{\beta \in \mathcal{B}}|\bar L(\beta) - L(\beta) | = o_p(1)$. The last equality holds because $\bar L(\beta) - L(\beta)$ can be seen as an empirical process over the Glivenko-Cantelli class $\mathcal{G}$, where $\mathcal{G} = \{w\xrightarrow{} \mathcal{A}^2_u(\beta,w), \beta \in \mathcal{B}\}$. The Glivenko-Cantelli property follows by Theorem 2.7.11 of \citet{Vaart1996} and the fact that the map $\beta \xrightarrow{}\mathcal{A}_u(\beta,w)$ is Lipschitz and bounded over $\mathcal{B}$ thanks to Assumption \ref{conditionE} (c), and therefore the same holds for the map $\beta \xrightarrow{}\mathcal{A}^2_u(\beta,w)$ is so. \hfill $\Box$\\

\begin{lemma}
\label{lemma:theo1lemma2}
     Under Assumption \ref{conditionR}, \ref{conditionC}  and \ref{conditionE}, if $\beta_0(u)$ is globally identified, define 
    $$
    \hat V = n^{-1} \sum_{i=1}^n \Bigg\{ \frac{\partial \mathcal{A}_u}{\partial \beta}(\beta_0,W_i) \left[ \hat{\mathcal{A}}_u(\beta_0,W_i) - \mathcal{A}_u(\beta_0,W_i) \ \right]\Bigg\},
    $$
    then there exists a Gaussian random variable $V$ such that $n^{1/2}\hat V\stackrel{d}{\rightarrow} V$, for $n\xrightarrow{} \infty$,  where $\stackrel{d}{\rightarrow}$ denotes convergence in distribution. 
\end{lemma}
\textbf{Proof:} We now show that $\hat V$ can be written as a U-statistic plus a negligible term.  Similarly as before, we add and subtract the term $\tilde{\mathcal{A}}_u$ defined in \eqref{eqb4} in the expression of $\hat V$: 
$$
\hat V = n^{-1} \sum_{i=1}^n \Bigg\{ \frac{\partial \mathcal{A}_u}{\partial \beta}(\beta_0,W_i) \left[ \hat{\mathcal{A}}_u - \tilde{\mathcal{A}}_u + \tilde{\mathcal{A}}_u \right](\beta_0,W_i)\Bigg\},
$$
and we first consider separately the two differences $(\hat{\mathcal{A}}_u -\tilde{\mathcal{A}}_u)(\beta_0,W_i)$ and $(\tilde{\mathcal{A}}_u -\mathcal{A}_u)(\beta_0,W_i)$. For the first difference, we have
    \begin{align*}
      &(\hat{\mathcal{A}}_u -\tilde{\mathcal{A}}_u )(\beta_0,W_i) \\
      &= n^{-1}\sum_{j=1}^n \delta_i \mathds{1} \{Y_j\le\exp(Z^\top_i\beta_0) , W_j\le W_i\}\Bigg[\frac{1}{ \hat G(Y_j)} - \frac{1}{ G(Y_j)}\Bigg]\\
            &= n^{-1}\sum_{j=1}^n \delta_i \mathds{1} \{Y_j\le\exp(Z^\top_i\beta_0) , W_j\le W_i\}\Bigg[\frac{G(Y_j) - \hat G(Y_j) }{G(Y_j) \hat G(Y_j) }  - \frac{G(Y_j) - \hat G(Y_j) }{G(Y_j)^2 } + \frac{G(Y_j) - \hat G(Y_j) }{G(Y_j)^2 } \Bigg]\\
    &= n^{-1}\sum_{j=1}^n \delta_i \mathds{1} \{Y_j\le\exp(Z^\top_i\beta_0) , W_j\le W_i\}\Bigg[\frac{G(Y_j) - \hat G(Y_j) }{G(Y_j)^2 } \Bigg] + O_p(n^{-1}),
    \end{align*}
where in the last equality, we used the fact that \begin{align}\label{eq:KMnew}
    \sup\limits_{t\in[0,\tau]}\Bigg| G(t) - \hat G(t)\Bigg| = O_p(n^{-1/2}),
\end{align}
with 
$\tau = \inf\limits_{t\in \mathbb{R}}\{ \sup\limits_{z\in\mathcal{Z}, \beta \in \mathcal{B}}\exp(z^{\top}\beta)< t\} $. The property \eqref{eq:KMnew} is a consequence of Theorem 1.1 in \citet{gill1983large} and Assumption \ref{conditionE} (e), which implies $\tau<\min(\bar c,\bar t)$. 

Next, we consider $\rho_{kji} =\frac{ \delta_j \upsilon_k(Y_j) }{G(Y_j)^2 }   \mathds{1} \{Y_j\le\exp(Z^\top_j\beta_0) , W_j\le W_i\}$, where $\upsilon_k(t) = \xi(Y_k,1-\delta_k,t)$ with  $\xi(y,\delta,t)$ defined at page 456 of \citet{lo1986product}. We can write 
$$(\hat{\mathcal{A}}_u -\tilde{\mathcal{A}}_u )(\beta_0,W_i)=n^{-2}\sum_{k,j=1}^n \rho_{kji}+O_p(n^{-1}).$$

Regarding the second difference $(\tilde{\mathcal{A}}_u -\mathcal{A}_u)(\beta_0,W_i)$, it also takes the following form:
\begin{align*}
  (\tilde{\mathcal{A}}_u -\mathcal{A}_u)(\beta_0,W_i) = n^{-1} \sum_{j=1}^n  \psi_{ji}, 
\end{align*}
where
\begin{align*}
   \psi_{ji} &= \frac{\delta_j}{G(Y_j)} \mathds{1} \{Y_j\le\exp(Z^\top_j\beta_0) , W_j\le W_i\} -  u\mathds{1}(W_j\le W_i) \\
    &\quad - \mathbb{E}\left[\frac{\delta}{G(Y)}
\mathds{1} \{Y\le \exp(Z^\top\beta_0), W\le W_i\}\right] +uP(W\le W_i). 
\end{align*}
Finally, define 
\begin{align*}
\phi_i &= \frac{\partial \mathcal{A}_u}{\partial \beta}(\beta_0,W_i). 
\end{align*}
Thus, we can rewrite $\hat V$ as 
\begin{align*}
    \hat V &= n^{-1} \sum_{i=1}^n \phi_i \Bigg[ n^{-1}\sum_{j=1}^n \left(\psi_{ji} + n^{-1} \sum_{k=1}^n\rho_{kji}\right)  \Bigg] + O_p(n^{-1})\\
     &= n^{-3} \sum_{i,j,k=1}^n \phi_i(\psi_{ji} + \rho_{kji})+ O_p(n^{-1}).
\end{align*}
We now show that $\hat V$ can be also written as a symmetric sum plus a negligible term. Precisely, define the following quantities:
\begin{align*}
\tilde\xi_{ijk} &= \phi_i(\psi_{ji} + \rho_{kji});\\
\xi_{kji} &=  \frac{1}{6}(\tilde\xi_{ijk}+\tilde\xi_{ikj} + \tilde\xi_{jik}+\tilde\xi_{jki} + \tilde\xi_{kji}+\tilde\xi_{kij}).
\end{align*}
Then,  $\hat V$ can be written as follows
\begin{align*}
    \hat V = n^{-3} \sum_{i,j,k=1}^n  \tilde\xi_{kji} + O_p(n^{-1}),
\end{align*}
where the last summation is symmetric in $i,j,k$ as aimed.
Since the terms involved are bounded, it is easy to see that $\hat V = \hat U +o_p(n^{-1/2}) +  O_p(n^{-1}),$
where 
$$
\hat U = n^{-3}  \sum_{i<j<k\in\{1,\dots,n\}}\xi_{ijk},
$$
which is a U-statistic. Applying now the Central Limit Theorem for U-statistics  (Theorem 1.1 in \citet{bose2018u}), we obtain the result. \hfill$\square$

\begin{lemma} \label{lemma:lipBound}
Under Assumptions \ref{conditionR} and \ref{conditionE}, there exists $M>0$ such that
$$
\mathbb{E}\Big[\left|\mathds{1}\{ T\le \exp(Z^\top\beta)\} - \mathds{1}\{T\le \exp(Z^\top\beta_0)\} \right| \Big]\le M \|\beta -\beta_0\|,
$$
for all $\beta\in\{\beta\in \mathbb{R}^K: \|\beta - \beta_0\|<\epsilon\}$, where $\epsilon$ is defined in Assumption \ref{conditionE} (g).
\end{lemma}
\textbf{Proof:} 
First, notice that 
\begin{align*}&\left|\mathds{1}\{ T\le \exp(Z^\top\beta)\} - \mathds{1}\{T\le \exp(Z^\top\beta_0)\} \right|\\
&=\mathds{1}\{ T\in[\exp(Z^\top\beta)\wedge\exp(Z^\top\beta_0),\exp(Z^\top\beta)\vee\exp(Z^\top\beta_0) ] \}.
 \end{align*}

Then, define
\begin{align*}
    L(z,\beta,\beta_0) =\exp(z^\top\beta)\wedge\exp(z^\top\beta_0); &\quad  U(z,\beta,\beta_0) = \exp(z^\top\beta)\vee\exp(z^\top\beta_0); \\
    l(z,\beta,\beta_0) = z^\top\beta\wedge z^\top\beta_0;&\quad u(z,\beta,\beta_0) = z^\top\beta\vee z^\top\beta_0.
\end{align*} Consider the following equality:
\begin{align*}
    \mathbb{E}\Big[\mathds{1}\big\{ T\in \big[L(Z,\beta,\beta_0),U(Z,\beta,\beta_0)\big] \big\} \Big] = \mathbb{E}\Bigg[\int_{L(Z,\beta,\beta_0)}^{U(Z,\beta,\beta_0) } f_{T|Z}(t|Z)dt \Bigg],
\end{align*}
and apply, consecutively, the integral mean value theorem and the mean value theorem for real functions, to obtain 
\begin{align*}
    \mathbb{E}\Bigg[\int_{L(Z,\beta,\beta_0)}^{U(Z,\beta,\beta_0) } f_{T|Z}(t|Z)dy\Bigg] &=     \mathbb{E}\left[\big\{U(Z,\beta,\beta_0)- L(Z,\beta,\beta_0)\big\} \big|f_{T|Z}(\theta_1(Z,\beta,\beta_0)|Z)\big|\right]\\
    &= \mathbb{E}\left[\big|Z^\top(\beta-\beta_0)\big|\exp(\theta_2(Z,\beta,\beta_0)) \big|f_{T|Z}(\theta_1(Z,\beta,\beta_0)|Z)\big|\right],
\end{align*}
where $\theta_1(Z,\beta,\beta_0)\in \big(L(Z,\beta,\beta_0),U(Z,\beta,\beta_0)\big)$ and 
$\theta_2(Z,\beta,\beta_0)\in \big(l(Z,\beta,\beta_0),u(Z,\beta,\beta_0)\big)$. Thanks to Assumption \ref{conditionE} (f), the density $f_{T|Z}$ is bounded. Moreover, thanks to Assumption \ref{conditionE} (g),  $\|z^\top(\beta-\beta_0)\exp(\theta_2(z,\beta,\beta_0))\|\le b(z)$ for every $z\in\mathcal{Z}$ and $\beta \in \{\beta\in \mathbb{R}^K: \|\beta - \beta_0\|<\epsilon\}$, where  $\epsilon$ and $b(z)$ are defined in Assumption \ref{conditionE} (g). Since we have $\mathbb{E}\big[b(Z)\big]<\infty$, this leads to
\begin{equation*}
     \mathbb{E}\Bigg[\int_{L(Z,\beta,\beta_0)}^{U(Z,\beta,\beta_0) } f_{T|Z}(t|Z)dy\Bigg] \le M\|\beta-\beta_0\|,
\end{equation*}
where $M$ is a constant, which is the assertion.  \hfill$\square$

\begin{lemma}
\label{lemma:VCclass}
The class of functions $\mathcal{G} = \{ (y,z)\in \mathbb{R}_{+}\times \mathcal{Z}\xrightarrow{} \mathds{1}\{y\le \exp(z^\top\beta)\} : \beta \in \mathbb{R}^K\}$ is VC. 
\end{lemma}
\textbf{Proof:} Consider the set $\mathcal{F} = \{f_{\alpha,\beta} : \alpha \in \mathbb{R},\beta \in \mathbb{R}^K \}$ where the function $f_{\alpha,\beta} : \mathbb{R}_{+}\times \mathcal{Z}\xrightarrow{} \mathbb{R}$ is defined as $f_{\alpha,\beta}(y,z) = z^\top\beta-\alpha\log(y)$, and note that $\mathcal{G}$ is a subset of  $\mathcal{H} = \{\mathds{1}\{f_{\alpha,\beta}\ge 0\}:f_{\alpha,\beta} \in \mathcal{F}\}$. Since $\mathcal{F}$ is a finite dimensional vector space, applying Lemma 9.6 in \citet{kosorok2008introduction}, $\mathcal{F}$ is VC. By  Lemma 9.9 (iii) in \citet{kosorok2008introduction}, we get that $ \{\{-f>0\},\ f\in\mathcal{F}\} $ is VC and, therefore, by Lemma 9.7 (i) in \citet{kosorok2008introduction} that $\{\{f\ge 0\},\ f\in\mathcal{F}\} $ is VC. Then, using the definition of a VC class, since $ \{ \{f_{0,\beta}\ge 0\},\ \beta\in\mathcal{B} \}\subset\{\{f\ge 0\},\ f\in\mathcal{F}\},$ we obtain that $\{ \{f_{0,\beta}\ge 0\},\ \beta\in\mathcal{B} \}$ is also a VC class. The statement of the Lemma is then a consequence of Lemma 9.8 in \citet{kosorok2008introduction}.\hfill$\square$

\subsection{Proof of Theorem \ref{theo:boot}}
The body of the proof is presented in the next section.  It relies on technical lemmas provided in Section \ref{subsec:lemmaboot} and it follows steps similar to that of the proof of Theorem \ref{theorem:2}. We denote by $\hat{\mathcal{A}}_{bu}(\beta,w)$ and $\hat L_b(\beta)$ the values of $\hat{\mathcal{A}}_u(\beta,w)$ and $\hat L(\beta)$ in the bootstrap sample. 
\subsubsection{Main Proof}
\textbf{Consistency} Using Lemma \ref{lemma:theo2lemma2} (c) and arguments similar to that of step 1 of Theorem \ref{theorem:2}, we obtain $\hat \beta_b - \beta_0 = o_p(1)$. \\

\noindent\textbf{Asymptotic expansion} Consider a sequence of $\beta\in\mathcal{B}$ such that $\beta\xrightarrow{}\beta_0$. Define $\bar{L}_b(\beta) = n^{-1} \sum_{i=1}^n \mathcal{A}^2_{u}(\beta,W_{bi})$
and write
\begin{align}
\hat L_b(\beta) - \bar L_b(\beta) \nonumber 
&=n^{-1}\sum_{i=1}^n\Bigg[ (\hat{\mathcal{A}}_{bu} -   {\mathcal{A}}_u)(\beta,W_{bi})\Bigg]^2 \\
&\quad +2n^{-1}\sum_{i=1}^n \mathcal{A}_u(\beta,W_{bi})\Bigg[ 
(\hat{\mathcal{A}}_{bu} - {\mathcal{A}}_u)(\beta,W_{bi}) 
\Bigg]. \label{boot:1}
\end{align}
We now consider separately the two terms on the right hand side of the last equation. Regarding the first, it can be rewritten as 
\begin{align*}
    n^{-1}\sum_{i=1}^n \Bigg[ (\hat{\mathcal{A}}_{bu}- {\mathcal{A}}_{u})(\beta,W_{bi}) \Bigg]^2 =
 n^{-1}\sum_{i=1}^n \hat{\mathcal{A}}_{bu}^2(\beta_0,W_{bi}) + I_{b1} +I_{b2},
\end{align*}
where the quantities $I_{b1}$ and $I_{b2}$ are
\begin{align*}
    I_{b1} &= n^{-1} \sum_{i=1}^n\Big[(\hat{\mathcal{A}}_{bu} - {\mathcal{A}}_u) (\beta,W_{bi})-\hat{\mathcal{A}}_{bu}(\beta_0,W_{bi})\Big]^2;\\
    I_{b2} &=2n^{-1}\sum_{i=1}^n\hat{\mathcal{A}}_{bu}(\beta_0,W_{bi})\Big[(\hat{\mathcal{A}}_{bu}-{\mathcal{A}}_u)(\beta,W_{bi})-\hat{\mathcal{A}}_{bu}(\beta_0,W_{bi})\Big].
\end{align*}
Now, by Lemma \ref{lemma:theo2lemma2} (a), it holds $|I_{b1}| = o_p(n^{-1})$. Moreover, using the inequality of Cauchy-Schwartz, we obtain 
\begin{align*}
    |I_{b2}| &\le 2\times \sqrt{n^{-1} \sum_{i=1}^n \hat{\mathcal{A}}_{bu}^2(\beta_0,W_{bi})}\\
    &\quad\quad \times \sqrt{n^{-1}\sum_{i=1}^n[(\hat{\mathcal{A}}_{bu}-{\mathcal{A}}_u)(\beta,W_{bi})-\hat{\mathcal{A}}_{bu}(\beta_0,W_{bi}) ]^2}.
\end{align*}
By Lemma \ref{lemma:theo2lemma2} (a) and (b), it also holds $I_{b2} = o_p(n^{-1})$ and we obtain
\begin{align}
    n^{-1}\sum_{i=1}^n \left[ \hat{\mathcal{A}}_{bu}(\beta,W_{bi}) - {\mathcal{A}}_u(\beta,W_{bi}) \right]^2 =
    n^{-1}\sum_{i=1}^n \hat{\mathcal{A}}_{bu}^2(\beta_0,W_{bi}) +o_p(n^{-1}). \label{boot:2}
\end{align}
Regarding the second term on the right hand side of Equation \eqref{boot:1}, we can use a Taylor expansion, Lemma \ref{lemma:theo2lemma2} (b) and the definition of $\hat V_b $ in Lemma \ref{lemma:theo2lemma3}, to justify the following equalities:
\begin{align}
&    2n^{-1}\sum_{i=1}^n \mathcal{A}_u(\beta,W_{bi})\left[ 
(\hat{\mathcal{A}}_{bu} - {\mathcal{A}}_u)(\beta,W_{bi}) 
\right] \nonumber\\ 
&\quad=2(\beta-\beta_0)^\top n^{-1}\sum_{i=1}^n \frac{\partial \mathcal{A}(\beta_0)(u,W_{bi})}{\partial \beta}\left[ 
(\hat{\mathcal{A}}_{bu} - {\mathcal{A}}_u)(\beta_0,W_{bi}) 
\right] + o_p(\|\beta - \beta_0\|^2) \nonumber\\
&\quad=2(\beta-\beta_0)^\top n^{-1/2}\hat V_b + o_p(n^{-1/2}\|\beta-\beta_0\| + \|\beta - \beta_0\|^2 ). \label{boot:3}
\end{align}
Next, equations \eqref{boot:1}, \eqref{boot:2}, \eqref{boot:3} yield
\begin{align*}
    \hat L_b(\beta) - \bar L_b(\beta)  
    &=n^{-1}\sum_{i=1}^n \hat{\mathcal{A}}_{bu}^2(\beta,W_{bi}) + 2(\beta-\beta_0)^\top n^{-1/2}\hat V_b \\
    &\quad + o_p(n^{-1/2}\|\beta-\beta_0\| + \|\beta - \beta_0\|^2 + n^{-1}).
\end{align*}
Using now a second-order Taylor expansion, we obtain
\begin{align*}
     &\hat L_b(\beta) - \hat L_b(\hat\beta) + o_p(n^{-1/2}\|\beta-\beta_0\| + \|\beta - \beta_0\|^2 + n^{-1})\\
     &= (\beta-\hat\beta)^\top n^{-1/2}\hat V_b + \frac{1}{2}(\beta -\beta_0)^\top\nabla^2\bar L(\beta_0)(\beta-\beta_0)-\frac{1}{2}(\hat\beta -\beta_0)^\top\nabla^2\bar L(\beta_0)(\hat\beta-\beta_0).
     \end{align*}
Adding and subtracting the term $(\beta -\hat\beta)^\top\nabla^2\bar L(\beta_0)(\hat\beta-\beta_0)$ and then using the equality  $  
\hat\beta -\beta_0 = -n^{-1/2}\nabla^2\bar L(\beta_0)^{-1}\hat V + o_p(n^{-1/2})$ from the proof of Theorem \ref{theorem:2}, we get
     \begin{align*}
     &\hat L_b(\beta) - \hat L_b(\hat\beta) + o_p(n^{-1/2}\|\beta-\beta_0\| + \|\beta - \beta_0\|^2 + n^{-1})\\
       &= (\beta-\hat\beta)^\top n^{-1/2}\hat V_b + \frac{1}{2}(\beta -\hat\beta)^\top\nabla^2\bar L(\beta_0)(\beta-\hat\beta)+(\beta -\hat\beta)^\top\nabla^2\bar L(\beta_0)(\hat\beta-\beta_0)\\
       &= (\beta-\hat\beta)^\top n^{-1/2}(\hat V_b - \hat V) + \frac{1}{2}(\beta -\hat\beta)^\top\nabla^2\bar L(\beta_0)(\beta-\hat\beta) + o_p(n^{-1/2}\|\beta - \hat \beta\|).
\end{align*}
Since $\hat\beta - \beta_0 = O_p(n^{-1/2})$, we obtain
\begin{align}
 \label{boot:4}
    \hat L_b(\beta) - \hat L_b(\hat \beta) &= (\beta-\hat \beta)^\top n^{-1/2}(\hat V_b -\hat V) +\frac{1}{2}(\beta-\hat\beta)\nabla^2\bar L(\beta_0)(\beta-\hat \beta)\nonumber\\
      &\quad\quad + o_p(n^{-1/2}\|\beta-\hat\beta\| + \|\beta - \hat\beta\|^2 + n^{-1}).
\end{align}
Since equation \eqref{boot:4} and \eqref{eq:taylorExpansion} are similar, the remainder is similar to that of Theorem \ref{theorem:2}. Hence, it is omitted. 

\subsubsection{Lemmas}\label{subsec:lemmaboot}
\begin{lemma}
\label{lemma:theo2lemma1}
Under Assumptions \ref{conditionR}, \ref{conditionC}  and \ref{conditionE}, if $\beta_0(u)$ is identified, then:
\setcounter{bean}{0}
\begin{list}
	{(\alph{bean})}{\usecounter{bean}}
	\item   $\sup_{ \beta \in \mathcal{B}, w\in \mathcal{W}} \left| (\hat{\mathcal{A}}_{bu} - \hat{\mathcal{A}}_u)(\beta,w) \right| = O_{P^*}(n^{-1/2})$.
	\item  $\lim\limits_{\epsilon \xrightarrow{}0  }\sup\limits_{\| \beta - \beta_0 \|\le \epsilon,w\in \mathcal{W}}  n^{1/2} \left| (\hat{\mathcal{A}}_{bu}- \hat{\mathcal{A}}_u)(\beta,w) - (\hat{\mathcal{A}}_{bu} -\hat{\mathcal{A}}_u)(\beta_0)(u,w)  \right| = o_{P^*}(1)$.
	\item  $\sup_{\beta \in \mathcal{B}}\left| \hat L_b(\beta) - \hat L(\beta)\right|=o_{P^*}(1)$.
\end{list}
\end{lemma}
\textbf{Proof:} The proof is similar to the one provided for Lemma \ref{lemma:theo1lemma1}. We report here the main differences. We replace the Donsker Theorem by the Donsker Theorem for bootstrap, see Theorem 3.6.3 of \citet{Vaart1996}. The uniform convergence for the bootstrap of the Kaplan-Meier estimator is obtained thanks to Theorem 1 in \citet{lo1986product}, the Donsker theorem for the bootstrap and the continuous mapping theorem. 
\begin{lemma} \label{lemma:theo2lemma2}
Under Assumption \ref{conditionR}, \ref{conditionC}  and \ref{conditionE}, if $\beta_0(u)$ is globally identified, then:
\setcounter{bean}{0}
\begin{list}
	{(\alph{bean})}{\usecounter{bean}}
	\item  $\sup_{w\in \mathcal{W}, \beta \in \mathcal{B}} | (\hat{\mathcal{A}}_{bu} - {\mathcal{A}}_{bu})(\beta,w) | = O_p(n^{-1/2})$.
	\item  $\lim\limits_{\epsilon \xrightarrow{}0  }\sup\limits_{\| \beta - \beta_0 \|\le \epsilon,w\in \mathcal{W}}  n^{1/2} | (\hat{\mathcal{A}}_{bu} - {\mathcal{A}}_u)(\beta,w) - \hat{\mathcal{A}}_{bu}(\beta_0,w)  | = o_{P^*}(1)$.
	\item  $\sup_{\beta \in \mathcal{B}}| \hat L_b(\beta) -  L(\beta)|=o_{P^*}(1)$.
\end{list}
\end{lemma}

\textbf{Proof:} The results follow from the triangle inequality, Lemma \ref{lemma:theo1lemma1}, Lemma \ref{lemma:theo2lemma1} and $\hat \beta - \beta_0 = o_p(1)$.  \hfill$\square$

\begin{lemma} \label{lemma:theo2lemma3}
Under Assumptions \ref{conditionR}, \ref{conditionC}  and \ref{conditionE}, if $\beta_0(u)$ is identified, define
    $$
    \hat V_b = n^{-1} \sum_{i=1}^n \Bigg\{ \frac{\partial \mathcal{A}_{u}}{\partial \beta}(\beta_0,W_{bi}) \left[ \hat{\mathcal{A}}_{bu}(\beta_0,W_{bi}) \right]\Bigg\},
    $$
    then  $n^{1/2}(\hat V_b - \hat V)\stackrel{d}{\rightarrow} V$, conditionally on the original sample in P-probability, where $V$ is specified in Lemma \ref{lemma:theo1lemma2}.
\end{lemma}
\textbf{Proof:} The proof is similar to the proof of Lemma \ref{lemma:theo1lemma2}. The only difference is that we use the bootstrap central limit theorem for U-statistics \citet{bickel1981asymptotic} instead of the central limit theorem for U-statistics. \hfill$\square$

\section{Further details on the empirical application}

\begin{figure}[htbp]
\includegraphics[width=0.8\textwidth, height=6cm]{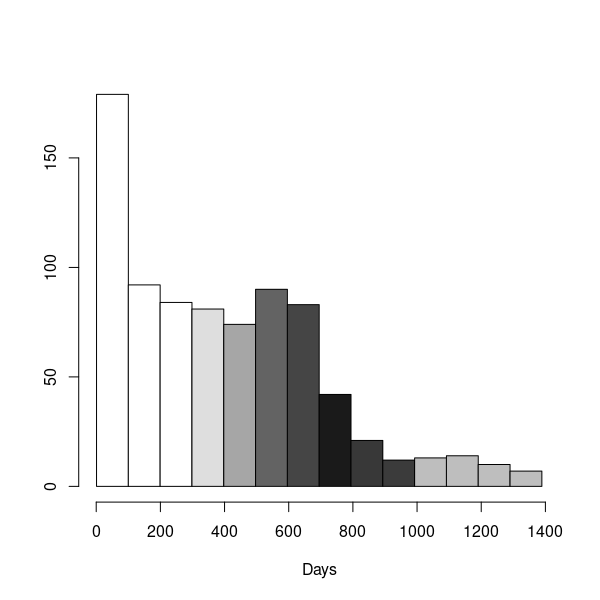}
\centering 
\caption{\label{fig:hist_job} Histogram of the values of $Y$ in the sample of non-white single mothers unemployed at the time of randomization. The darker the shade, the higher the proportion of censoring. }
\end{figure}
\begin{table}[htbp]
\caption{\label{table:tablenumemp2} Empirical application results}
\begin{center}
\begin{tabular}{|c|c|r|r|r|}
	\hline
	$u$                    & Component & \multicolumn{1}{c|}{Estimation} & \multicolumn{1}{c|}{CI 0.025} & \multicolumn{1}{c|}{CI 0.975} \\ \hline
	\multirow{3}{*}{0.1} & intercept & 4.701                           & 4.132                       & 5.732                        \\ \cline{2-5} 
	& treatment & -0.236                          & -0.979                      & 0.306                        \\ \cline{2-5} 
	& age       & -0.035                          & -0.079                      & -0.004                       \\ \hline
	\multirow{3}{*}{0.2} & intercept & 5.930                           & 5.424                       & 7.095                        \\ \cline{2-5} 
	& treatment & -0.571                          & -1.466                      & -0.174                       \\ \cline{2-5} 
	& age       & -0.047                          & -0.089                      & -0.016                       \\ \hline
	\multirow{3}{*}{0.3} & intercept & 6.861                           & 5.900                       & 7.569                        \\ \cline{2-5} 
	& treatment & -0.927                          & -1.375                      & 0.085                        \\ \cline{2-5} 
	& age       & -0.052                          & -0.091                      & -0.018                       \\ \hline
	\multirow{3}{*}{0.4} & intercept & 6.792                           & 6.282                       & 7.558                        \\ \cline{2-5} 
	& treatment & -0.486                          & -0.921                      & -0.070                       \\ \cline{2-5} 
	& age       & -0.044                          & -0.079                      & -0.018                       \\ \hline
	\multirow{3}{*}{0.5} & intercept & 7.083                           & 6.587                       & 8.585                        \\ \cline{2-5} 
	& treatment & -0.701                          & -1.597                      & -0.188                       \\ \cline{2-5} 
	& age       & -0.036                          & -0.071                      & -0.006                       \\ \hline
	\multirow{3}{*}{0.6} & intercept & 8.280                           & 7.793                       & 8.934                        \\ \cline{2-5} 
	& treatment & -1.520                          & -2.008                      & -1.054                       \\ \cline{2-5} 
	& age       & -0.038                          & -0.062                      & 0.001                        \\ \hline
\end{tabular}
\end{center}
\footnotesize
\renewcommand{\baselineskip}{11pt}
\textbf{Note:} 
 Estimates and bounds of bootstrap 95\% confidence intervals for the variables \textit{intercept}, \textit{treatment} and \textit{age} for the quantiles $u\in\{0.1,...,0.6\}$.
\end{table}

\end{document}